\documentclass[longauth]{aa}  
\usepackage{graphicx}
\usepackage{txfonts}
\usepackage{hyperref}
\usepackage{xcolor}
\usepackage{multirow}
\usepackage{multicol}
\usepackage{threeparttable}
\usepackage{siunitx}
\usepackage[capitalise]{cleveref}
\usepackage{mhchem}

\newcommand{\vsini}{$V \sin i$}

\newcommand{\teff}{$T_{\rm eff}$}
\newcommand{\logg}{log\,{\it g$_\star$}}

\newcommand{\feh}{[Fe/H]}
\newcommand{\cah}{[Ca/H]}
\newcommand{\sih}{[Si/H]}
\newcommand{\mgh}{[Mg/H]}
\newcommand{\nah}{[Na/H]}

\newcommand{\kms}{km\,s$^{-1}$}
\newcommand{\ms}{m~s$^{-1}$}

\newcommand{\gc}{g~cm$^{-3}$}

\newcommand{\Lsun}{$L_{\odot}$}                          
\newcommand{\Msun}{$M_{\odot}$}
\newcommand{\Rsun}{$R_{\odot}$}
\newcommand{\mearth}{$M_{\oplus}$}

\newcommand{\rearth}{$R_{\oplus}$}

\newcommand{\Porb}{$P_{\rm orb}$} 
\newcommand{\Prot}{$P_{\rm rot}$} 
\newcommand{\mstar}{$M_\star$}
\newcommand{\rstar}{$R_\star$}
\newcommand{\lstar}{$L_\star$}
\newcommand{\rhostar}{$\rho_{\mathrm{*}}$}
\newcommand{\rhoplanet}{$\rho_\mathrm{p}$}
\newcommand{\rplanet}{$R_{\mathrm{p}}$}
\newcommand{\mplanet}{$M_{\mathrm{p}}$}
\newcommand{\Teq}{$T_{\rm eq}$}

\newcommand{\citla}{\texttt{citlalicue}}
\newcommand{\george}{\texttt{george}}
\newcommand{\pyt}{\texttt{pytransit}}
\newcommand{\pyan}{\texttt{pyaneti}}

\newcommand{\Tzerob}[1][days]   {$8545.7767 _{ - 0.0023 } ^ { + 0.0031 }$~#1} 
\newcommand{\Pb}[1][days]   {$4.884765 _{ - 2.4e-5 } ^ { + 1.9e-5 }$~#1} 
\newcommand{\esinb}[1][ ]   {$-0.08 \pm 0.19 $~#1} 
\newcommand{\ecosb}[1][ ]   {$0.01 _{ - 0.16 } ^ { + 0.15 }$~#1} 
\newcommand{\bb}[1][ ]   {$0.29 _{ - 0.19 } ^ { + 0.20 }$~#1} 
\newcommand{\arb}[1][ ]   {$14.0 _{ - 0.88 } ^ { + 0.80 }$~#1} 
\newcommand{\rrb}[1][ ]   {$0.01925 _{ - 0.00085 } ^ { + 0.00079 }$~#1} 
\newcommand{\kb}[1][${\rm m\,s^{-1}}$]   {$2.23 \pm 0.26 $~#1} 
\newcommand{\mpb}[1][$M_{\oplus}$]   {$5.72 _{ - 0.68 } ^ { + 0.70 }$~#1} 
\newcommand{\rpb}[1][$R_{\oplus}$]   {$1.992 _{ - 0.090 } ^ { + 0.085 }$~#1} 
 
\newcommand{\eb}[1][ ]   {$0.046 _{ - 0.033 } ^ { + 0.056 }$~#1} 
\newcommand{\wb}[1][deg]   {$-53.2 _{ - 68.1 } ^ { + 158.9 }$~#1} 
 
\newcommand{\ib}[1][deg]   {$88.85 _{ - 0.82 } ^ { + 0.77 }$~#1} 
\newcommand{\ab}[1][AU]   {$0.0618 _{ - 0.0039 } ^ { + 0.0036 }$~#1}

\newcommand{\insolationb}[1][${\rm F_{\oplus}}$]   {$207.1 _{ - 23.5 } ^ { + 29.9 }$~#1} 
\newcommand{\tsmb}[1][ ]   {$46.29 _{ - 7.47 } ^ { + 9.26 }$~#1}

\newcommand{\Teqb}[1][K]   {$1055.8 _{ - 31.3 } ^ { + 36.2 }$~#1} 
\newcommand{\ttotb}[1][hours]   {$2.61 _{ - 0.10 } ^ { + 0.15 }$~#1} 
\newcommand{\tfulb}[1][hours]   {$2.50 _{ - 0.11 } ^ { + 0.15 }$~#1} 
\newcommand{\tegb}[1][hours]   {$0.0542 _{ - 0.0047 } ^ { + 0.0112 }$~#1} 
 
\newcommand{\denpb}[1][${\rm g\,cm^{-3}}$]   {$3.98 _{ - 0.66 } ^ { + 0.77 }$~#1} 
\newcommand{\grapb}[1][${\rm cm\,s^{-2}}$]   {$1752 _{ - 321 } ^ { + 340 }$~#1} 
 
\newcommand{\jspb}[1][ ]   {$20.61 _{ - 2.68 } ^ { + 2.78 }$~#1} 
\newcommand{\qone}[1][]   {$0.31 _{ - 0.22 } ^ { + 0.39 }$~#1} 
\newcommand{\qtwo}[1][]   {$0.31 _{ - 0.22 } ^ { + 0.35 }$~#1}

\newcommand{\harps}[1][${\rm km\,s^{-1}}$]   {$-23.81711 _{ - 0.00061 } ^ { + 0.00083 }$~#1} 
\newcommand{\FWHM}[1][${\rm km\,s^{-1}}$]   {$6.9541 _{ - 0.0083 } ^ { + 0.0105 }$~#1} 
\newcommand{\jharps}[1][${\rm m\,s^{-1}}$]   {$1.08 _{ - 0.21 } ^ { + 0.23 }$~#1} 
\newcommand{\jFWHM}[1][${\rm m\,s^{-1}}$]   {$3.24 _{ - 0.34 } ^ { + 0.40 }$~#1} 
\newcommand{\jtr}[1][]   {$597.8 _{ - 5.3 } ^ { + 5.4 }$~#1} 
\newcommand{\jArvc}[1][\ms]   {$0.00081 _{ - 0.00056 } ^ { + 0.00178 }$~#1} 
\newcommand{\jArvr}[1][\msd]   {$0.0187 _{ - 0.0083 } ^ { + 0.0278 }$~#1} 
\newcommand{\jAfwhm}[1][\ms]   {$0.0116 _{ - 0.0051 } ^ { + 0.0160 }$~#1} 
\newcommand{\lambdae}[1][days]   {$162.4 _{ - 41.2 } ^ { + 27.3 }$~#1} 
\newcommand{\lambdap}[1][]   {$0.57 _{ - 0.15 } ^ { + 0.29 }$~#1} 
\newcommand{\PGP}[1][days]   {$25.48 _{ - 0.14 } ^ { + 0.15 }$~#1}

\newcommand{\smassariadne}[1][]{$0.956^{+0.050}_{-0.026}$}   
\newcommand{\smassariadnegrav}[1][]{$0.97^{+0.11}_{-0.10}$}  
\newcommand{\smassparam}[1][]{$0.931 \pm 0.035$}   

\newcommand{\sradiusariadne}[1][]{$0.949^{+0.008}_{-0.012}$}   
\newcommand{\sradiusparam}[1][]{$0.944\pm 0.024$}  
\newcommand{\srgaia}[1][]{$0.950^{+0.050}_{-0.031}$}  

\newcommand{\srhoariadne}[1][]{$1.58 \pm 0.19$}   
\newcommand{\srhoparam}[1][]{$1.56 \pm 0.13$}   

\newcommand{\steffariadne}[1][]{$5639 \pm 31$} 
\newcommand{\sloggariadne}[1][]{$4.47 \pm 0.05$} 
\newcommand{\sfehariadne}[1][]{$-0.05 \pm 0.05$} 
\newcommand{\Lumariadne}[1][]{$0.82 \pm 0.02$} 
\newcommand{\Avariadne}[1][]{$0.01 \pm 0.02$} 

\newcommand{\steffsme}[1][]{$5585 \pm 60$} 
\newcommand{\sloggsme}{$4.47 \pm 0.05$} 
\newcommand{\scahsme}[1][]{$-0.01 \pm 0.05$} 
\newcommand{\sfehsme}[1][]{$-0.04 \pm 0.05$} 
\newcommand{\snasme}[1][]{$+0.04 \pm 0.05$} 
\newcommand{\smghsme}[1][]{$+0.03 \pm 0.05$} 
\newcommand{\ssihsme}[1][]{$+0.02 \pm 0.05$} 
\newcommand{\svsini}[1][]{$2.2 \pm 0.7$} 
\newcommand{\svmic}[1][]{$1.0$} 
\newcommand{\svmac}[1][]{$2.8$} 
\newcommand{\velsme}[1][$\mathrm{km\,s^{-1}}$]{$xxx \pm x$} 

\newcommand{\steffspechmatch}[1][]{$5554 \pm 110$} 
\newcommand{\sloggspechmatch}[1][]{$4.30 \pm 0.12$} 
\newcommand{\sfehspechmatch}[1][]{$-0.09 \pm 0.09$} 
\newcommand{\sagespechmatch}[1][]{$7.1 \pm 1.5$} 

\newcommand{\sloggparam}[1][]{$4.43 \pm 0.03$} 

\newcommand{\distancegaia}{$75.27\pm 0.07$} 
\newcommand{\parallaxgaia}{$13.2847\pm0.0127$} 
\newcommand{\velgaia}[1][$\mathrm{km\,s^{-1}}$]{$-23.9 \pm 0.15$} 
\newcommand{\pmra}{$27.528 \pm 0.007$} 
\newcommand{\pmdec}{$19.524 \pm 0.012$} 

\newcommand{\spectraltype}{G6\,V} 
\newcommand{\spectraltyperadius}{0.95} 
\newcommand{\spectraltypemass}{0.97} 

\newcommand{\ageparam}{$6.2\pm 3.6$} 
\newcommand{\ageariadne}[1][]{$4.4^{+1.5}_{-3.1}$} 

\begin{document}

\title{TOI-733~b: A planet in the small-planet radius valley orbiting a Sun-like star
\thanks{Based on observations made with the ESO-3.6 m telescope at La Silla Observatory under programme 106.21TJ.001.}\thanks{Table~\ref{all_rv.tex} only available in electronic form
at the CDS via \url{https://cdsarc.cds.unistra.fr/cgi-bin/qcat?J/A+A/}.}}
\titlerunning{TOI-733}

\author{Iskra~Y.~Georgieva\inst{\ref{OSO}} 
\and
Carina~M.~Persson\inst{\ref{OSO}} 
\and
Elisa~Goffo\inst{\ref{Torino},\ref{Tautenburg}} 
\and
Lorena~Acu\~na\inst{\ref{Marseille}, \ref{MaxPlanck}} 
\and
Artyom~Aguichine\inst{\ref{Marseille}, \ref{SantaCruz}} 
\and
Luisa~M.~Serrano\inst{\ref{Torino}}
\and
Kristine~W.~F.~Lam\inst{\ref{DLR}}
\and
Davide~Gandolfi\inst{\ref{Torino}} 
\and
Karen~A.~Collins\inst{\ref{HarvardSmithsonian}} 
\and
Steven~B.~Howell\inst{\ref{NASAames}}
\and
Fei~Dai\inst{\ref{Pasadena}, \ref{Fei2}, \ref{Fei3}}
\and
Malcolm~Fridlund\inst{\ref{OSO}, \ref{Leiden}} 
\and
Judith~Korth\inst{\ref{Lund},\ref{Chalmers}}
\and
Magali~Deleuil\inst{\ref{Marseille}}
\and
Oscar~Barrag\' an\inst{\ref{Oxford}}
\and
William~D.~Cochran\inst{\ref{McDonald}}
\and
Szil\' ard~Csizmadia\inst{\ref{DLR}}
\and
Hans~J.~Deeg\inst{\ref{IAC}, \ref{LaLaguna}}
\and
Eike~Guenther\inst{\ref{Tautenburg}}
\and
Artie~P.~Hatzes\inst{\ref{Tautenburg}}
\and
Jon~M.~Jenkins\inst{\ref{NASAames}}
\and
John~Livingston\inst{\ref{AstrobiologycentreTokyo},\ref{SOKENDAI}, \ref{NAOTokyo}} 
\and
Rafael~Luque\inst{\ref{UniversityChicago}} 
\and
Olivier~Mousis\inst{\ref{Marseille}}
\and
Hannah~L.~M.~Osborne\inst{\ref{MullardVincent}}
\and
Enric~Palle\inst{\ref{IAC}}
\and
Seth~Redfield\inst{\ref{Wesleyan}}
\and
Vincent~Van~Eylen\inst{\ref{MullardVincent}}
\and
Joseph~D.~Twicken\inst{\ref{NASAames}, \ref{SETI}}
\and
Joshua~N.~Winn\inst{\ref{Princeton}}
\and
Ahlam~Alqasim \inst{\ref{MullardVincent}}
\and
Kevin~I.~Collins\inst{\ref{GeorgeMason}}
\and
Crystal~L.~Gnilka\inst{\ref{NASAames}, \ref{NASAExo}}
\and
David~W.~Latham\inst{\ref{HarvardSmithsonian}}
\and
Hannah~M.~Lewis\inst{\ref{STScI}}
\and
Howard~M.~Relles\inst{\ref{HarvardSmithsonian}}
\and
George~R.~Ricker\inst{\ref{MIT Kavli}}
\and
Pamela~Rowden\inst{\ref{PamelaRowden}}
\and
Sara~Seager\inst{\ref{SaraSeager1}, \ref{Avi}, \ref{SaraSeager3}}
\and
Avi~Shporer\inst{\ref{Avi}}
\and
Thiam-Guan~Tan\inst{\ref{Perth}}
\and
Andrew~Vanderburg \inst{\ref{Vanderburg1}, \ref{Vanderburg2}, \ref{Fei3}}
\and 
Roland~Vanderspek\inst{\ref{MIT Kavli}}
}

\offprints{iskra.georgieva@chalmers.se}
    \institute{Department of Space, Earth and Environment, Chalmers University of Technology, Onsala Space Observatory, SE-439 92 Onsala, Sweden. \label{OSO}
       \email{\url{iskra.georgieva@chalmers.se}}
\and Dipartimento di Fisica, Universita degli Studi di Torino, via Pietro Giuria 1, I-10125, Torino, Italy \label{Torino}  
\and  Th\"uringer Landessternwarte Tautenburg, Sternwarte 5, 07778 Tautenburg, Germany  \label{Tautenburg}
\and Aix Marseille Universit\'e, Institut Origines, CNRS, CNES, LAM, Marseille, France  \label{Marseille} 
\and Max-Planck-Institut f\"ur Astronomie, K\"onigstuhl 17, D-69117 Heidelberg, Germany \label{MaxPlanck} 
\and Department of Astronomy \& Astrophysics, University of California, Santa Cruz, CA 95064, USA \label{SantaCruz} 
\and Institute of Planetary Research, German Aerospace Center (DLR), Rutherfordstrasse 2, D-12489 Berlin, Germany \label{DLR} %
\and Center for Astrophysics \textbar \ Harvard \& Smithsonian, 60 Garden Street, Cambridge, MA 02138, USA \label{HarvardSmithsonian}
\and NASA Ames Research Center, Moffett Field, CA 94035, USA \label{NASAames}
\and Division of Geological and Planetary Sciences, 1200 E California Blvd, Pasadena, CA, 91125, USA \label{Pasadena}   
\and Department of Astronomy, California Institute of Technology, Pasadena, CA 91125, USA \label{Fei2} 
\and NASA Sagan Fellow \label{Fei3} 
\and Leiden Observatory, University of Leiden, PO Box 9513, 2300 RA, Leiden, The Netherlands \label{Leiden} 
\and Lund Observatory, Division of Astrophysics, Department of Physics, Lund University, Box 43, SE-221 00 Lund, Sweden
\label{Lund} 
\and Department of Space, Earth and Environment, Chalmers University of Technology, Chalmersplatsen 4, 412 96 Gothenburg, Sweden \label{Chalmers}
\and Sub-department of Astrophysics, Department of Physics, University of Oxford, Oxford, OX1 3RH, UK \label{Oxford}
\and  McDonald Observatory and Center for Planetary Systems Habitability, The University of Texas, Austin Texas USA \label{McDonald}  
\and Instituto de Astrof\' isica de Canarias, C. Via Lactea S/N, E-38205 La Laguna, Tenerife, Spain\label{IAC}
\and Universidad de La Laguna, Dept. de Astrof\'isica, E-38206 La Laguna, Tenerife, Spain \label{LaLaguna}  
\and Astrobiology Center, 2-21-1 Osawa, Mitaka, Tokyo 181-8588, Japan  \label{AstrobiologycentreTokyo}    
\and Department of Astronomical Science, The Graduate University for Advanced Studies (SOKENDAI), 2-21-1 Osawa, Mitaka, Tokyo, Japan  \label{SOKENDAI}
\and National Astronomical Observatory of Japan, 2-21-1 Osawa, Mitaka, Tokyo 181-8588, Japan  \label{NAOTokyo}
\and Department of Astronomy \& Astrophysics, University of Chicago, Chicago, IL 60637, USA  \label{UniversityChicago}
\and Mullard Space Science Laboratory, University College London, Holmbury St Mary, Dorking, Surrey RH5 6NT, UK \label{MullardVincent}
\and Astronomy Department and Van Vleck Observatory, Wesleyan University, Middletown, CT 06459, USA \label{Wesleyan}
\and SETI Institute, Mountain View, CA 94043, USA. \label{SETI}
\and Department of Astrophysical Sciences, Princeton University, Princeton, NJ 08544, USA \label{Princeton} 
\and George Mason University, 4400 University Drive, Fairfax, VA, 22030 USA. \label{GeorgeMason} 
\and Space Telescope Science Institute, 3700 San Martin Drive, Baltimore, MD, 21218, USA \label{STScI}
\and NASA Exoplanet Science Institute, Caltech/IPAC, Mail Code 100-22, 1200 E. California Blvd., Pasadena, CA 91125, USA \label{NASAExo} 
\and MIT Kavli Institute for Astrophysics and Space Research \& MIT Physics Department \label{MIT Kavli} 
\and Royal Astronomical Society, Burlington House, Piccadilly, London W1J 0BQ, UK \label{PamelaRowden}
\and Department of Earth, Atmospheric, and Planetary Sciences, Massachusetts Institute of Technology, Cambridge, MA 02139, USA \label{SaraSeager1}
\and Department of Physics and Kavli Institute for Astrophysics and Space Research, Massachusetts Institute of Technology, Cambridge, MA 02139, USA \label{Avi}
\and Department of Aeronautics and Astronautics, Massachusetts Institute of Technology, Cambridge, MA 02139, USA \label{SaraSeager3}
\and Perth Exoplanet Survey Telescope, Perth, Western Australia \label{Perth}
\and Department of Astronomy, The University of Wisconsin–Madison, Madison, WI 53706, USA \label{Vanderburg1}
\and Department of Astronomy, The University of Texas at Austin, Austin, TX 78712, USA \label{Vanderburg2}
}

\date{Received Date Month YYYY; accepted Date Month YYYY}

 
  \abstract
  {We report the discovery of a hot (\Teq\ $\approx$ 1055~K) planet in the small-planet radius valley that transits the Sun-like star TOI-733.\ It was discovered as part of the \mbox{KESPRINT} follow-up program of TESS planets carried out with the HARPS spectrograph. TESS photometry from sectors 9 and 36 yields an orbital period of \Porb\ = \Pb\ and a radius of \rplanet\ = \rpb. Multi-dimensional Gaussian process modelling of the radial velocity measurements from HARPS and activity indicators  gives a semi-amplitude of $K$ = \kb, translating into a planet mass of \mplanet\ = \mpb. These parameters imply that the planet is of moderate density ($\rho_\mathrm{p}$\ = \denpb) and place it in the transition region between rocky and volatile-rich planets with H/He-dominated envelopes on the mass-radius diagram. Combining these with stellar parameters and abundances, we calculated planet interior and atmosphere models, which in turn suggest that TOI-733~b has a volatile-enriched, most likely secondary outer envelope, and may represent a highly irradiated ocean world. This is one of only a few such planets around G-type stars that are well characterised.
  }
   
%

   \maketitle
%

\section{Introduction} \label{Section: intro}

The end of the last millennium saw the addition of a new field to astronomy: the field of exoplanets. Since 2000, thousands of planets have been discovered by Convection, Rotation and planetary Transits  \citep[\textit{CoRoT};][]{Baglin2006}, the Kepler space telescope \citep[\textit{Kepler}; ][]{2010Sci...327..977B,2014PASP..126..398H}, and the currently operating Transiting Exoplanet Survey Satellite \citep[TESS; ][]{2015JATIS...1a4003R}. The latter has followed in the footsteps of the indispensable \textit{Kepler} and provided data that have led to the confirmation of about 300\footnote{\href{NASA Exoplanet Archive}{\url{https://exoplanetarchive.ipac.caltech.edu/}}. Accessed 16 January 2023.} exoplanets so far, with thousands more to be confirmed in the years to come.

Space transit surveys, in particular \textit{Kepler}, have facilitated the confirmation of the theoretically predicted \citep{2013ApJ...775..105O, 2013ApJ...776....2L} and observationally demonstrated \citep{2017AJ....154..109F, VanEylen2018} small-planet radius gap. This region, evident in planet radius versus orbital period (equally versus planet equilibrium emperature or stellar irradiation), is characterised by a dearth of planets with radii near 1.8~\rearth~\citep{2017AJ....154..109F, VanEylen2018}. On the lower radius side are the super-Earths, which are rocky with or without thin secondary envelopes. On the larger radius side are the so-called mini-Neptunes with typically slightly larger cores and more significant H/He-dominated envelopes. The radius valley is the manifestation of the separation between the two. 
The origin of the radius gap has been investigated in detail, and two main theories have arisen: atmospheric photoevaporation resulting from intense stellar irradiation \citep{2013ApJ...775..105O, 2013ApJ...776....2L}, and core-powered mass loss, that is, atmospheric mass loss driven by leftover heat from formation that escapes from the core
 \citep{2018MNRAS.476..759G, 2019MNRAS.487...24G}.

The planet mass, however, is the crucial parameter that, when combined with the radius, allows us to begin to characterise the detected planets. The relative faintness of the stellar targets in \textit{Kepler}'s primary mission has unfortunately made the determination of this fundamental property difficult. Currently, one of the most high-profile ambiguities in exoplanet science is the composition degeneracy \citep{Valencia07, Zeng2016, Zeng2019} of the in between planets found in the radius valley. 
It is important to characterise planets with precisely determined radius and mass in this region because this is a key ingredient in the recipe for breaking the degeneracy. By mapping out and disentangling the structure of these interesting objects, we may be able to uncover new pathways to planet formation and evolution. Modern high-precision spectrographs work well, and the yield of planetary mass is expected to be sufficient to allow population studies that are not limited only to radius \citep[e.g. ][]{2022A&A...668A.178K}.

In addition to the commonly assumed composition of a silicate mantle surrounding an iron core with a H/He envelope on top, a possibility in systems that are not young (a few billion years old) is that a planet between 1.6--2.5~\rearth\ can be helium enhanced \citep{2022NatAs.tmp..248M}.
Transition region planets have also been hypothesised to be water worlds, however, or to feature a significant H$_2$O content, a possible volatile atmosphere \citep[e.g. ][]{Zeng2019, 2021ApJ...923..247Z, Mousis20}. Recently, \citet{Luque22} showed that small planets around M~dwarfs are likely to be water worlds, whose existence can be explained via type I migration from beyond the snow line. The authors suggested that their conclusions might be extended to solar-type stars. The recent discussions and analyses of the systems Kepler-138 presented by \citet{2023NatAs...7..206P}, K2-3 by \citet{2022AJ....164..172D}, and of TOI-1695 by \citet{Cherubim_2023} also indicate that water-dominated planets might be more prevalent than previously thought, even at super-Earth radii. 

Furthermore, we cannot distinguish with
current population studies whether photoevaporation or core-powered mass loss is the mechansim that sculpts the radius gap, as shown by \citet{2021MNRAS.508.5886R}. One of the key ingredients to determine the mechanism, as they point out, is obtaining high-accuracy planet radii and stellar host masses in systems in which the planets reside in  or close to the radius gap. Thus, improving our understanding of the origins and histories of these planets is a crucial part of the way to crystallise the widely studied phenomenon of atmospheric mass loss. 

In this paper, we present the discovery and characterisation of a planet inside the small-planet radius valley, TOI-733~b (TIC 106402532), which was discovered by TESS in 2019. We show that its possible compositions make it a particularly interesting and important planet that can serve as a stepping stone to showing that a population of water worlds also exists around Sun-like stars, as well as to reduce the uncertainty surrounding the aforementioned problems.  

In Sect.~\ref{Section: obs} we present all space- and ground-based observations performed on TOI-733 and the data analysis. Section~\ref{Section: Stellar modelling} describes our stellar modelling, and Sect.~\ref{Section: transit and RV modelling} summarises our transit and multi-dimensional Gaussian process (multi-GP) modelling. In Sect.~\ref{Section: disco} we present the placement of TOI-733~b in the small planet population, as well as our interior and atmospheric modelling. Our conclusions are laid out in Sect.~\ref{Section: tada}.

\begin{figure*}
  \centering
  \includegraphics[width=\textwidth]{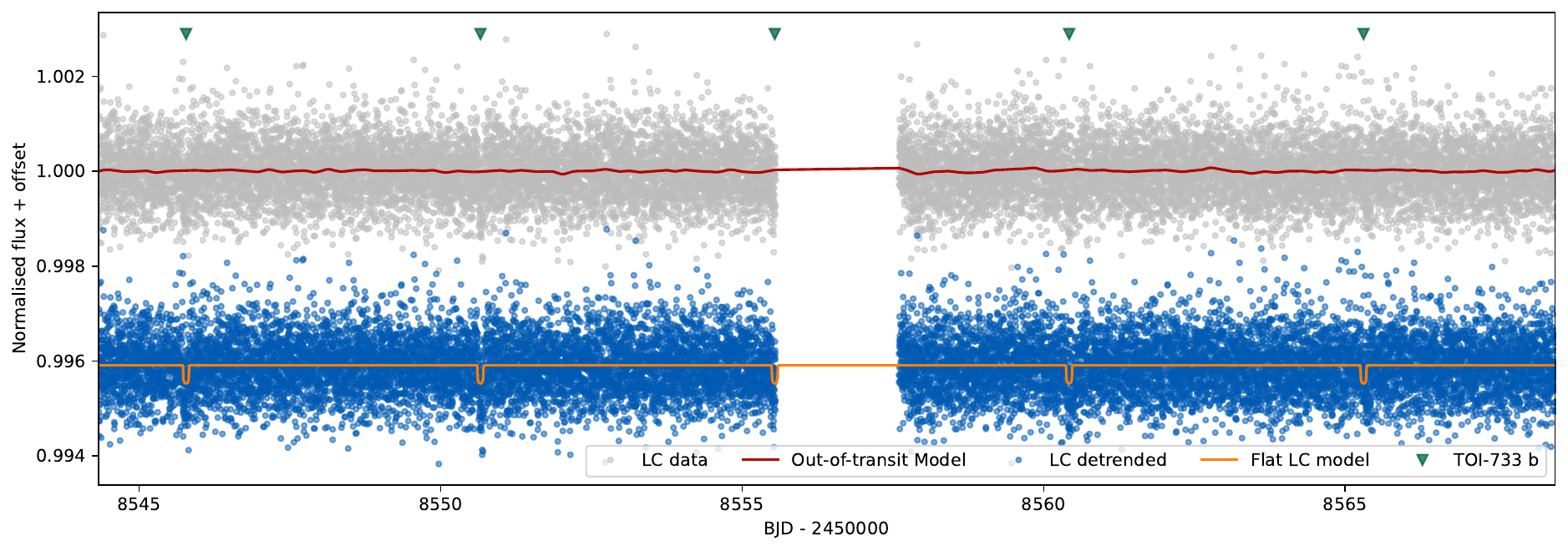}
  \includegraphics[width=1\textwidth]{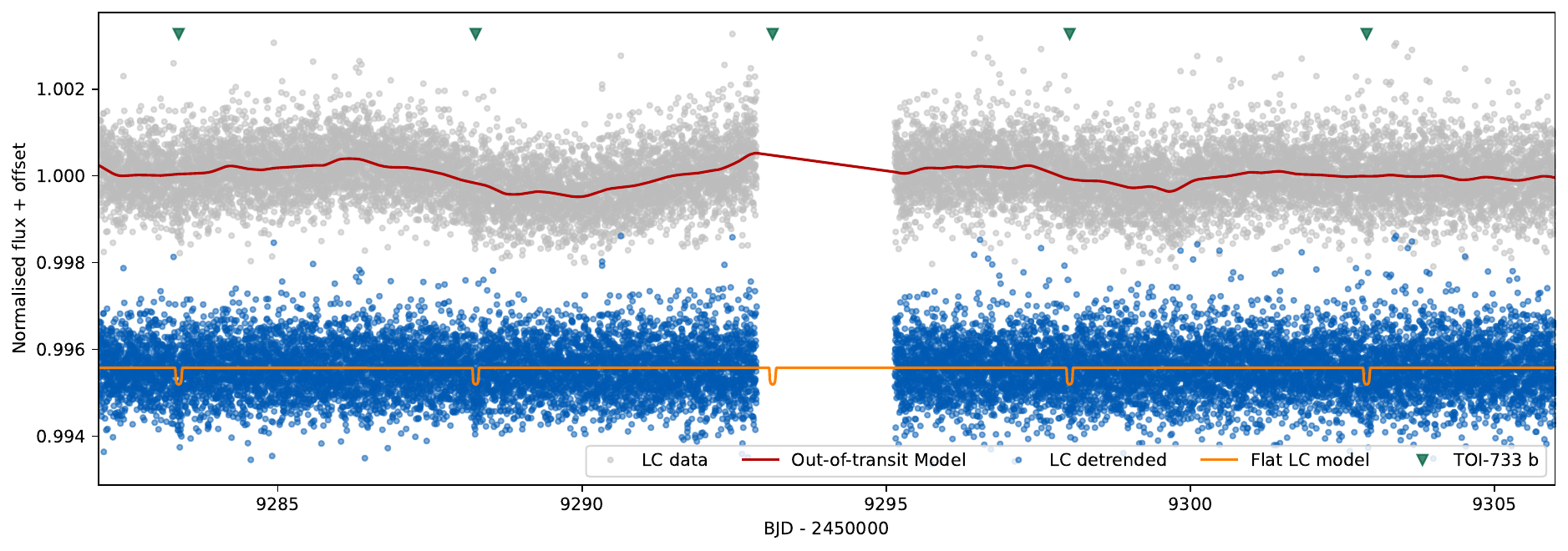}
    \caption{Light curves (LC) with a two-minute cadence in sectors 9 (top) and 36 (bottom) of TESS PDCSAP are plotted in grey. The locations of the individual transits of TOI-733~b are marked by green triangles. The GP-inferred model for the out-of-transit data is marked by the red curve. The vertically offset blue dots show the resulting detrended light curve, and the orange fit with transits is overplotted.}.
    \label{Figure: citla}
\end{figure*}

\section{TOI-733 space- and ground-based observations} \label{Section: obs}  

\begin{table}[!t]
\caption{Basic parameters for TOI-733.}
\begin{center}
 \resizebox{0.9\columnwidth}{!}{%
\begin{tabular}{lll} 
\hline\hline
     \noalign{\smallskip}
Parameter    & Value   \\
\noalign{\smallskip}
\hline
\noalign{\smallskip}
\multicolumn{2}{l}{\emph{Main Identifiers}} \\
\noalign{\smallskip}
TIC  & 106402532 \\
2MASS &  J10373820-4053179 \\
WISE &  J103738.22-405317.7 \\
TYC & 7714-00657-1 \\
UCAC4 & 246-045192 \\
Gaia   &        5392409372314518656     \\
\noalign{\smallskip}
\hline
\noalign{\smallskip}
\multicolumn{2}{l}{\emph{Equatorial coordinates (epoch 2015.5)}} \\
\noalign{\smallskip} 
R.A. $(J2000.0)$ & $10\fh37\fm 38\fs24$     \\
Dec. $(J2000.0)$ & -40$\fdg 53\farcm17\farcs73$    \\
\noalign{\smallskip}
\hline
\noalign{\smallskip}
\multicolumn{2}{l}{\emph{Magnitudes}} \\
TESS & $8.8411 \pm 0.0060$ &  \\
Johnson $ B$ &   $10.4900\pm0.0167$     \\
Johnson $V$  & $9.435\pm0.019$      \\
 $G \tablefootmark{a}$  &   $9.2966\pm0.0001$ \\
 $G_{RP}\tablefootmark{a}$      &   $8.7875\pm0.0007$ \\
$G_{BP} \tablefootmark{a}$      &   $9.6335\pm0.0007$ \\
$J$  &   $8.220\pm 0.026$     \\
$H$    &   $7.943\pm0.040$      \\
$K$   &     $7.845\pm0.024$   \\
WISE $W1$   &     $7.780\pm0.027$   \\
WISE $W2$   &     $7.851\pm0.020$   \\
\noalign{\smallskip}
\hline
\noalign{\smallskip}  
Parallax$\tablefootmark{a}$  (mas) &\parallaxgaia    \\ 
Distance$\tablefootmark{a}$  (pc) &\distancegaia    \\
$\mu_{RA}\tablefootmark{a}$ (mas~yr$^{-1}$) & \pmra   \\   
$\mu_{Dec}\tablefootmark{a}$ (mas~yr$^{-1}$) & \pmdec   \\  
\noalign{\smallskip}
\hline
\hline
\noalign{\smallskip}
\rstar$\tablefootmark{b}$ (\Rsun)  & \sradiusariadne \\
\mstar$\tablefootmark{b}$  (\Msun) & \smassariadne \\
\rhostar$\tablefootmark{b}$ (\gc)  & \srhoariadne \\
\lstar$\tablefootmark{b}$ (\Lsun)  & \Lumariadne \\ 
Age$\tablefootmark{b}$ (Gyr) & \ageariadne \\
\teff$\tablefootmark{b}$ (K)  & \steffsme \\
\logg$\tablefootmark{b}$    &  \sloggsme \\
\feh$\tablefootmark{b}$    & \sfehsme \\
\cah$\tablefootmark{b}$    & \scahsme\\
\mgh$\tablefootmark{b}$   & \smghsme \\
\nah$\tablefootmark{b}$    & \snasme \\
\sih$\tablefootmark{b}$    & \ssihsme \\
\vsini$\tablefootmark{b}$ (\kms) &\svsini\\
    \noalign{\smallskip} \noalign{\smallskip}
\hline 
\end{tabular}
}
\tablefoot{
\tablefoottext{a}{Gaia DR3.} 
\tablefoottext{b}{This work (Sect.~\ref{Section: Stellar modelling}).} 
}
\end{center}
\label{Table: Star basic parameters}
\end{table} 

To confirm the planetary nature of the candidate TOI-733.01, we relied upon space-based light-curve photometry from TESS, follow-up ground-based photometry from Las Cumbres Observatory Global Telescope (LCOGT), speckle imaging from the Zorro instrument at the 8m Gemini South telescope, and spectroscopy by the HARPS spectrograph at the 3.6m telescope at La Silla observatory. The target identifiers and coordinates together with other relevant stellar parameters are listed in Table~\ref{Table: Star basic parameters}.

\subsection{Photometry from TESS} \label{photo}  

The field of sector 9 was observed during the first TESS cycle between 2019 February 28 UT and 2019 March 25 UT, and of sector 36  in the third cycle, between 2021 March 07 UT and 2021 April 01 UT. TOI-733 (TIC~106402532) was observed by camera 2, CCD 2, in the nominal two-minute cadence in both sectors. 

The data were processed in the TESS Science Processing Operations Center \citep[SPOC, ][]{jenkins2016} at NASA Ames Research Center. SPOC conducted a transit search of the light curve in sector 9 on 2019 April 25 and of the light curve in sector 36 on 2021 April 14 with an adaptive noise-compensating matched filter \citep{Jenkins02, Jenkins10, Jenkins2020}. The search produced a threshold crossing event (TCE) with a period of 4.887~d, for which an initial limb-darkened transit model was fitted \citep{Li2019} and a suite of diagnostic tests were conducted to help determine the planetary nature of the signal. The results of these tests can be found in the Data Validation Reports \citep[DVR,][]{Twicken2018}, which are available for download via the Mikulski Archive for Space Telescopes (MAST)\footnote{\url{https://archive.stsci.edu/}} and the {\tt EXOFOP-TESS} website\footnote{\url{https://exofop.ipac.caltech.edu/tess/target.php?id=106402532}}. The TESS Science Office (TSO) reviewed the vetting report and issued an alert for TOI 733.01 on 2019 June 6 \citep{Guerrero2021}.
The reports for the two sectors show no concerning traits regarding any contaminating sources in the aperture of the SPOC pipeline, which is generated for the production of simple aperture photometry
\citep[SAP, ][]{Twicken10, Morris20}.

In the absence of these potential complications and according to common practice, we downloaded the presearch data conditioning (PDCSAP) \citep{Smith2012, Stumpe2012, Stumpe2014} light curves from MAST and proceeded to use them for the transit analysis and light-curve modelling (\ref{Section: transit and RV modelling}).

To detrend the light curves, we used a Gaussian process (GP) type detrending with the code \href{https://github.com/oscaribv/citlalicue}{\citla}\footnote{\url{https://github.com/oscaribv/citlalicue}}, which is a wrapper of \href{https://github.com/dfm/george}\george\footnote{\url{https://github.com/dfm/george}}\ \citep{Mackey2014, Ambi2016} and  \href{https://github.com/hpparvi/PyTransit}\pyt\footnote{\url{https://github.com/hpparvi/PyTransit}}\ \citep{Parviainen2015}. This Python package fits a model to the out-of-transit data using likelihood maximisation to account for the stellar variability \citep[for more details, see e.g. ][]{Oscar-letter, Carina22}. The same procedure was applied to the data from both sectors. Figure~\ref{Figure: citla} shows the PDCSAP light curve in sector 36, the GP model, and the resulting detrended light curve. For the joint modelling (Sect.~\ref{Section: transit and RV modelling}), we only used the cutout transits and not the entire light curve to speed up the computation.

\subsection{Ground-based light-curve follow-up}
\label{Karen}  

The TESS pixel scale is $\sim 21\arcsec$ pixel$^{-1}$ , and photometric apertures typically extend out to roughly 1 arcminute, which generally results in the blending of multiple stars in the TESS aperture. To attempt to determine the true source of the TESS detection, we conducted ground-based photometric follow-up observations of the field around TOI-733 as part of the {\tt TESS} follow-up observing program\footnote{\url{https://tess.mit.edu/followup}} sub-group 1 \citep[TFOP;][]{collins:2019}. If the event detected in the TESS data were indeed on-target, the shallow SPOC-reported depth of $\sim 400$ ppm would not generally be detectable in ground-based observations. Instead, we slightly saturated TOI-733 to enable the extraction of nearby fainter star light curves to attempt to rule out or identify nearby eclipsing binaries (NEBs) as potential sources of the TESS detection.

We observed a predicted transit window of TOI-733.01 in the Sloan $i'$ band using the LCOGT \citep{Brown:2013} 1.0\,m network node at Cerro Tololo Inter-American Observatory (CTIO) on 2019 August 18 UT. The 1\,m telescopes are equipped with $4096\times4096$ SINISTRO cameras having an image scale of $0\farcs389$ per pixel, resulting in a $26\arcmin\times26\arcmin$ field of view. The images were calibrated by the standard LCOGT {\tt BANZAI} pipeline \citep{McCully:2018}, and photometric data were extracted using {\tt AstroImageJ} \citep{Collins:2017}.

We observed a second predicted transit of TOI-733.01 from the Perth Exoplanet Survey Telescope (PEST) near Perth, Australia. The 0.3 m telescope was equipped with a $1530\times1020$ SBIG ST-8XME camera with an image scale of 1$\farcs$2 pixel$^{-1}$ , resulting in a $31\arcmin\times21\arcmin$ field of view. A custom pipeline based on {\tt C-Munipack}\footnote{\url{http://c-munipack.sourceforge.net}} was used to calibrate the images and extract the differential photometry.

We scheduled full transit observations using the initial SPOC TESS sector 9 nominal ephemeris (P = 4.88651\,d, T0 = 1545.7732\,BTJD). A later SPOC sector 9 and 36 multi-year ephemeris (P = $4.88478\pm0.00002$\,d, T0 = $1545.7755\pm0.0015$\,BTJD) showed that our follow-up observations missed the revised predicted ingress, but covered the egress window with more than $\pm7\sigma$ timing uncertainty coverage. The multi-year SPOC centroid shift results limit the source to within $\sim30\arcsec$ of TOI-733 ($3\sigma)$. We therefore focused our NEB search on the nine known Gaia DR3 and TICv8 stars within $60\arcsec$ of TOI-733 that are bright enough in the TESS band to produce the TESS detection.

We calculated the root mean square (RMS) over the full duration of the raw light curve after normalising it to a mean value of 1.0. We repeated this for each of the nine nearby star light curves (binned in five-minute bins) and found that the RMS values of the LCOGT light curve are lower by more than a factor of 5 than the expected NEB depth in each respective star, except for the $3\arcsec$ neighbor TIC 865377947 and the $16\arcsec$ neighbor TIC 106402536. The TIC 865377947 photometric aperture is strongly blended with the much brighter target star TOI-733, and TIC 106402536 is contaminated with a TOI-733 diffraction spike that contains strong photometric systematics from the saturated target star. Although NEB signals cannot be ruled out in TIC 865377947 and TIC 106402536, we find that NEB signals are ruled out in the remaining seven nearby stars. In addition, the PEST light curve of TIC 106402536 excludes an NEB egress at a level of $3\times\rm{RMS}$. We then visually inspected the light curve of each neighboring star to ensure that there was no obvious deep eclipse-like signal. Through a process of elimination, we find that the TESS signal must occur in TOI-733 or in the $3\arcsec$ neighbor TIC 865377947, relative to known Gaia DR3 and TICv8 stars. Our follow-up light curves are available on the {\tt EXOFOP-TESS} website.

 \begin{figure}[!ht]
 \centering
  \resizebox{\hsize}{!}
            {\includegraphics{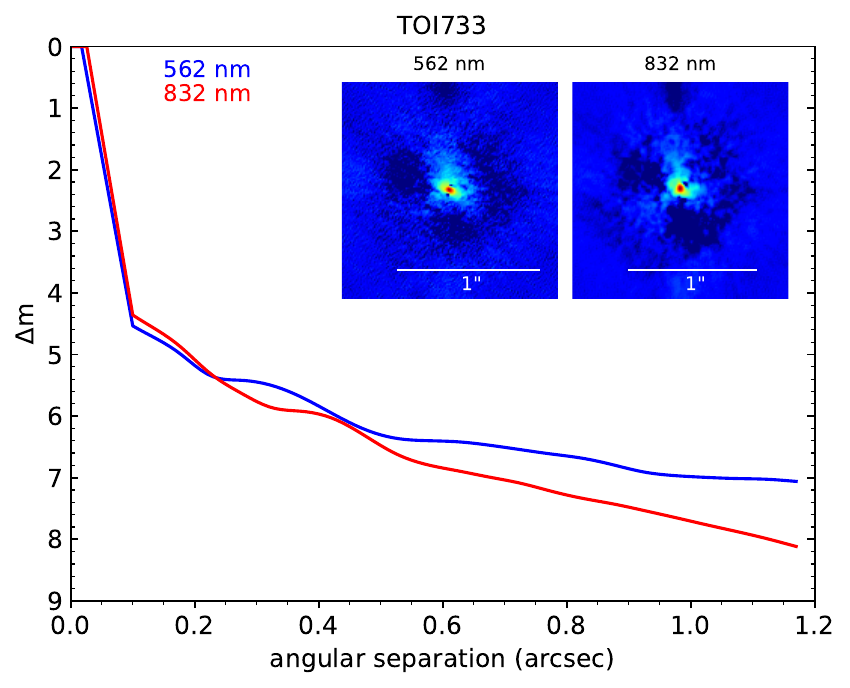}}
   \caption{5 $\sigma$ sensitivity curve resulting from the speckle imaging by Gemini South/Zorro. The reconstructed image shows that no bright companions are detected within 1.2\arcsec.}
      \label{Figure: speckle}
 \end{figure}

\subsection{Speckle imaging from Gemini-South/Zorro} \label{speckle} 

If an exoplanet host star has a spatially close companion, this companion (bound or line of sight) can create a false-positive transit signal if it is an eclipsing binary (EB), for example. Third-light flux from a close companion star can lead to an underestimated planetary radius if it is not accounted for in the transit model \citep{ciardi2015} and can even cause non-detections of small planets that reside within the same exoplanetary system \citep{2021AJ....162...75L}. The discovery of close, bound companion stars provides crucial information toward our understanding of exoplanetary formation, dynamics, and evolution \citep{2021AJ....161..164H}. Thus, to search for close-in bound companions that are unresolved in TESS or other ground-based follow-up observations, we obtained high-resolution speckle imaging observations of TOI-733.

TOI-733 was observed on 2020 March 15 UT using the Zorro speckle instrument on the Gemini South 8 m telescope \citep{2021FrASS...8..138S}. Zorro provides simultaneous speckle imaging in two bands (562nm and 832 nm). The output data products include a reconstructed image with robust contrast limits on companion detections. Three sets of 1000 X 0.06 sec exposures were collected and subjected to a Fourier analysis in our standard reduction pipeline \citep[see][]{howell2011}. Figure~\ref{Figure: speckle} shows our final contrast curves and the two reconstructed speckle images. We find that TOI-733 is a single star and has no companion brighter than 5-8 magnitudes (0.1\arcsec\ to 1.0\arcsec) below that of the target star from the diffraction limit (20 mas) out to 1.2\arcsec. At the distance of TOI-733 (d=75 pc), these angular limits correspond to spatial limits of  1.5 to 90 au.

\subsection{Spectroscopy and frequency analysis} \label{sect: freq}  

We observed TOI-733 with the High Accuracy Radial velocity Planet Searcher \citep[HARPS; ][]{Mayor03} spectrograph mounted at the ESO 3.6~m telescope of La Silla Observatory in Chile. We obtained a total of 74 high-resolution (R~$\approx$~115 000, $\lambda$~$\in$~378– 691~nm) spectra between 17 February  and 8 June 2022 UT as part of our large observing program 106.21TJ.001 (PI: Gandolfi). The exposure time varied between 1200 and 1800 seconds, depending on weather conditions and observing schedule constraints, leading to a signal-to-noise ratio (S/N) per pixel at 550 nm between 45 and 109. We used the second fibre of the instrument to simultaneously observe a Fabry-Perot interferometer and trace possible nightly instrumental drifts \citep{Wildi10, Wildi11}. The HARPS data were reduced using the dedicated data reduction software \citep[DRS; ][]{2007A&A...468.1115L} available at the telescope. For each spectrum, the DRS also provides the full width at half maximum (FWHM) and the bisector inverse slope (BIS) of the cross-correlation function (CCF). We also extracted additional activity indicators and spectral diagnostics, namely H$\alpha$, the S-index, the differential line width (dLW), and the chromatic index (crx) using the codes {\tt{serval}} \citep{Zechmeister2018} and TERRA \citep{2012ApJS..200...15A}. A snippet of the data is shown in Table~\ref{all_rv.tex}.

\begin{figure}
\centering
\includegraphics[width=\linewidth]{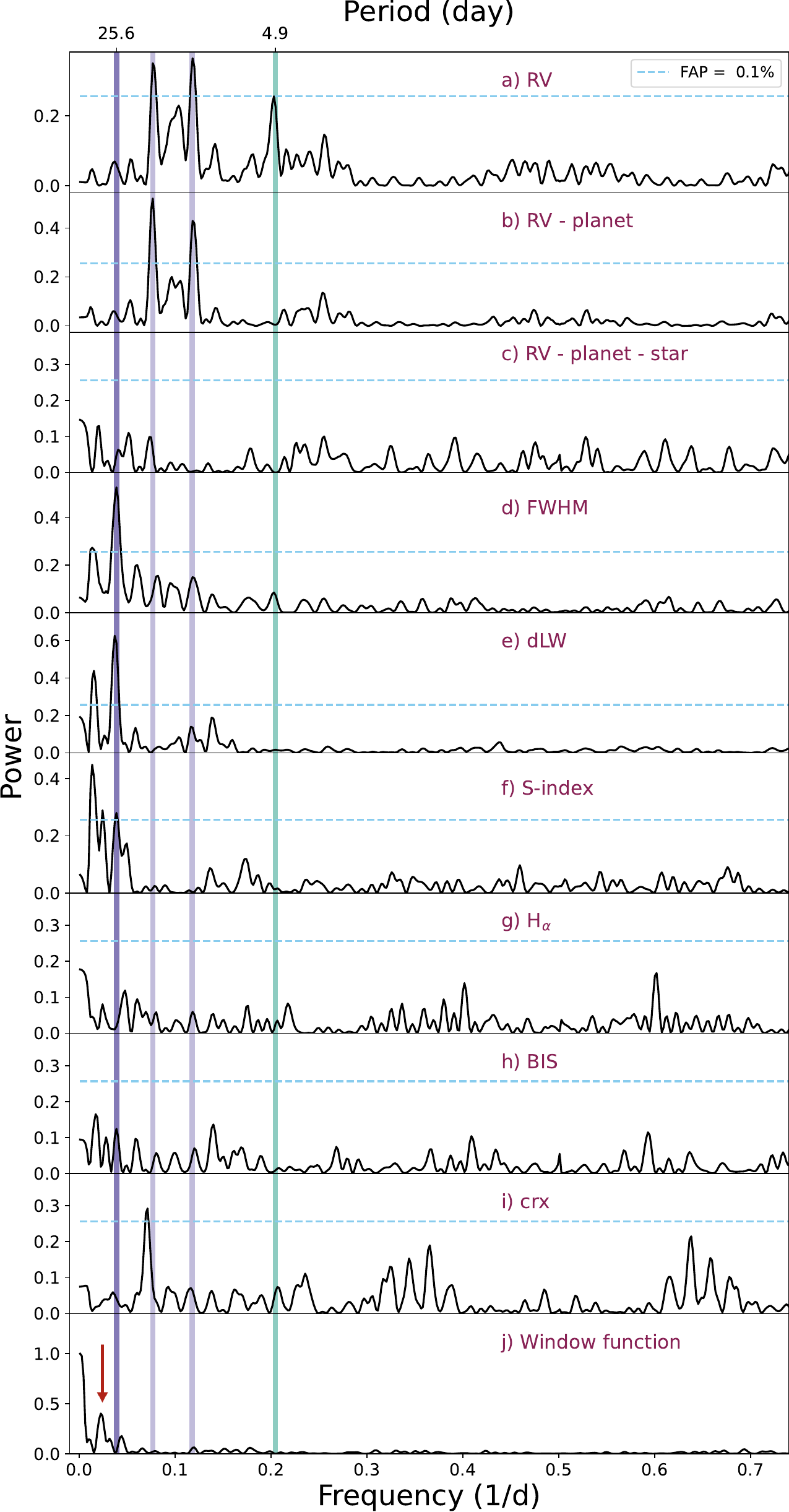}
    \caption{Generalized Lomb-Scargle periodograms of the spectroscopic data for TOI-733. From top to bottom, the top three panels correspond to the DRS RVs, to the RV residuals after fitting a sinusoid at the detected orbital period for TOI-733.01 (marked as solid vertical teal line), and to residuals after fitting the final multi-GP model presented in Sect.~\ref{Section: transit and RV modelling}. 
    The following panels show periodograms of spectral activity indicators, ending with the window function at the bottom, as annotated in each panel.
    The solid purple lines represent the frequency of the GP signal, which is particularly visible in the FWHM and dLW, while its first two harmonics in semi-transparent purple are well pronounced in the RVs. A peak at the first harmonic is also visible in panel i). The horizontal blue line shows the 0.1\,\% FAP level. 
\label{fig: GLS}}
\end{figure}

As a first step into investigating the TOI-733 spectroscopic data, we performed a frequency analysis to search for significant signals as potential signatures of orbiting planets and/or stellar activity. Figure~\ref{fig: GLS} shows the generalised Lomb Scargle \citep[GLS;][]{Zech09} periodograms of the HARPS RV data as extracted by the DRS pipeline, as well as common activity indicators from the DRS, {\tt{serval,}} and TERRA. 
We considered a signal to be significant if its false-alarm probability \citep[FAP;][]{1997A&A...320..831K} was lower than 0.1~\%. We used the bootstrap method to estimate the FAP, denoted here by a blue horizontal line in all but the bottom panel. 

The periodogram of the DRS RVs (upper panel) shows two significant (FAP~<~0.1~\%) peaks at 0.078~day$^{-1}$ and 0.118~day$^{-1}$  (semi-transparent purple), which correspond to periods of about 12.8 and 8.5 d, respectively. We note a third peak at the transit signal of TOI-733.01 (0.205~day$^{-1}$, teal vertical line). While this peak does not cross the FAP~=~0.1~\% threshold that we required to consider it significant, we can use our prior knowledge of the frequency of the transit signal to estimate the probability that noise might produce a peak at the orbital frequency of the transit signal whose power exceeds the orbserved power of TOI-733.01. Following the method described in \citet{2019dmde.book.....H}, we computed the GLS periodogram of 10$^{5}$ simulated data sets obtained by randomly shuffling the RV measurements while keeping the observation time-stamps fixed. We found that none of the 10$^{5}$ periodgram trials displays a peak at 0.205~~day$^{-1}$ with a power greater than the observed one, implying an FAP~<~0.001~\%.

The second panel shows a periodogram of the RV residuals after subtracting the signal of the planet candidate. The latter two peaks remain undisturbed. The FWHM and dLW periodograms (panels d and e, respectively) clearly show a peak at $\sim$~0.04~day$^{-1}$ (25.6~days, solid purple line). Although not clearly identifiable in the activity indicators, subtracting a signal at this frequency from the FWHM causes the peak at 8.5~days to become apparent. When the 8.5-day signal is subtracted, the signal at 12.8~days can also be identified (Fig.~\ref{Figure: GLS_res}). All this shows that all three signals (25.6~days, 12.8~days, and 8.5~days) are present in the FWHM, which in turn allows us to attribute the latter two (leftmost peaks in the top RV panel) to the first two harmonics of the 25.6-day signal. We thus consider the latter to be the true rotation period of the star and point out that such a \Prot\ is consistent with \rstar\ and the \vsini\ estimated in Sect.~\ref{Section: Stellar modelling}. 

It is also worth noting that the S-index panel also displays the significance of \Prot\ estimated in this way, but it is less prominent than the highest peak in this panel, which is at 65.2~days. The bottom panel shows the periodogram of the window function. It shows a peak at a frequency equal to the frequency spacing between 1/25.6 and 1/50~day$^{-1}$, that is, 0.0237~day$^{-1}$ (red arrow in the bottom panel), pointing to the interpretation that the 65.2-day signal is an alias of the rotation frequency. 

Panel c) presents the periodogram of the RV residuals after the final model described in Sect.~\ref{Section: transit and RV modelling} is subtracted from the data. No more significant peaks are present in the data. Together with an RV jitter term of $\sim$1.1~\ms\ (Table~\ref{Table: Orbital and planetary parameters}), this shows that based on the gathered observations, there is no evidence that an additional planet orbits TOI-733.

\section{Stellar modelling} \label{Section: Stellar modelling}
For the spectroscopic modelling of TOI-733, we used two codes:  SpecMatch-Emp  \citep{2017ApJ...836...77Y}, and SME\footnote{\url{http://www.stsci.edu/~valenti/sme.html}} 
 \citep[Spectroscopy Made Easy;][]{vp96, pv2017}, version 5.2.2. The latter fits  observations to synthetic spectra computed with  atomic and molecular line data     from the VALD\footnote{\url{http://vald.astro.uu.se}} \citep{Ryabchikova2015} and stellar atmosphere grids     
  \citep[Atlas12;][]{Kurucz2013}.   
 SpecMatch-Emp is   an emperical 
 code that compares observations   to a dense 
 library of  very well characterised FGKM 
 stars. This software finds an  effective 
 temperature \teff = \steffspechmatch~K, 
 a   surface gravity \logg  = \sloggspechmatch,  and 
an iron   abundance \feh = \sfehspechmatch. These values were used as a first input to the more elaborate SME modelling  \citep[further details of  the SME modelling can be found in][]{2018A&A...618A..33P}. In short,  
we fitted one parameter at a time using spectral lines that are particularly sensitive to the fitted parameters. 
We fixed the 
micro-turbulent velocity $V_{\rm mic}$ 
to \svmic~km~s$^{-1}$ \citep{bruntt08}   
and the macro-turbulent velocity $V_{\rm mac}$  to  \svmac~km~s$^{-1}$ \citep{Doyle2014}.
Our final SME  model gives  
 \teff = \steffsme~K,    
 \feh = \sfehsme,
 \cah = \scahsme,
\sih = \ssihsme,
\mgh = \smghsme, 
\nah = \snasme, 
\logg = \sloggsme,   
and a projected rotational velocity 
$V \sin i_\star = $~\svsini~km~s$^{-1}$    in excellent 
agreement with Specmatch-emp. 
The SME modelling  points to a \spectraltype~star with typical mass and radius of \spectraltypemass~\Msun~and \spectraltyperadius~\Rsun, respectively. 

  \begin{figure}[!ht]
 \centering
  \resizebox{\hsize}{!}
            {\includegraphics{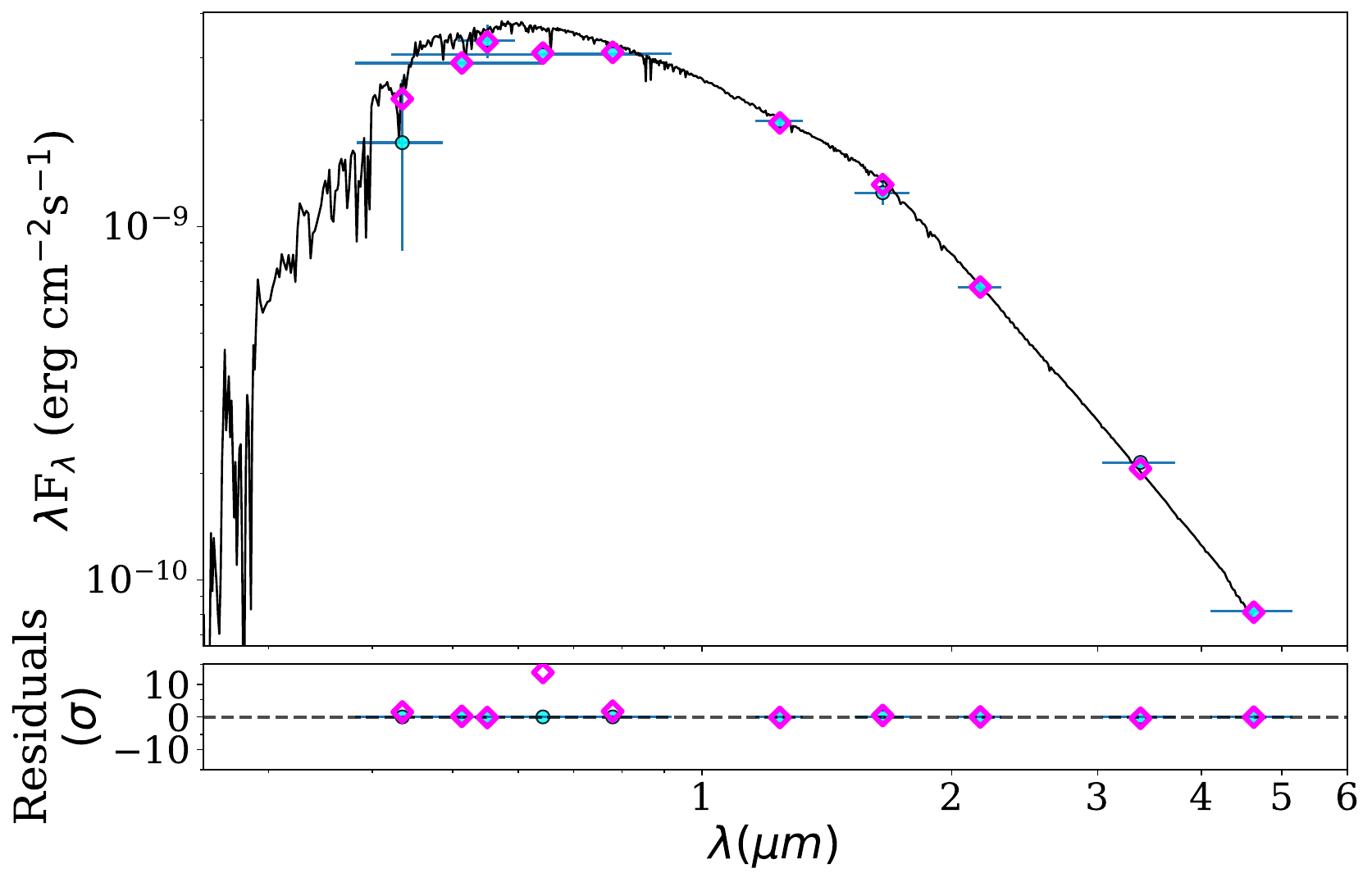}}
   \caption{Spectral energy distribution  of TOI-733 and the best-fit model 
   \citep{Castelli2004}. Magenta diamonds outline the  synthetic      photometry,  and the   observed 
photometry is shown with blue points.   We plot the 1~$\sigma$ uncertainties of the magnitudes (vertical error bars), and the effective width of the passbands is marked with horizontal bars. The residuals in the lower panel are normalised to the errors of the photometry.}
      \label{Figure: SED}
 \end{figure}

To  model the stellar 
radius, mass, and age, we used the 
python package   
ARIADNE\footnote{\url{https://github.com/jvines/astroARIADNE}}
\citep[][]{2022MNRAS.513.2719V}. 
With 
this software,  broadband photometry 
was fit to the spectral energy 
distribution (SED). We  fit 
the following bandpasses: Johnson $V$ 
and $B$  (APASS),
$G G_{\rm BP} G_{\rm RP}$   (DR3),    
$JHK_S$  ({\it 2MASS}), {\it WISE} 
W1-W2,    and 
the Gaia DR3 parallax. 
We set an upper limit of  $A_V$ based on the dust maps of \citet{1998ApJ...500..525S}. 
ARIADNE fits the photometric observations to the four 
atmospheric model grids {\tt {Phoenix~v2}} 
\citep{2013A&A...553A...6H}, {\tt {BtSettl}} 
\citep{2012RSPTA.370.2765A}, 
\citet{Castelli2004}, and 
\citet{1993yCat.6039....0K}, and 
computed the final radius   with Bayesian model
averaging. 
The  final stellar radius 
is  \rstar\ = \sradiusariadne~\rstar. We also find  a
luminosity of  \lstar\ = \Lumariadne~\Lsun,  and
an extinction that is consistent with
zero \mbox{($A_\mathrm{V} =$ \Avariadne)}. 
The stellar mass in   ARIADNE was 
interpolated from   the MIST
\citep{2016ApJ...823..102C} isochrones
and is found to be \mstar\ = \smassariadne~\mstar. When we combine the radius from ARIADNE and \logg~from SME, the gravitational mass is \smassariadnegrav~\Msun. The posteriors in the ARIADNE model for \teff, \feh, and \logg~agree very well with the priors taken from SME.

We checked the ARIADNE results with  {\tt {PARAM1.3}}\footnote{\url{http://stev.oapd.inaf.it/cgi-bin/param_1.3}} 
\citep{daSilva2006}. This software uses Bayesian computation and PARSEC isochrones with    \teff, \feh, the $V$ magnitude, 
and the Gaia DR3 parallax as priors. The results are in excellent agreement within 1~$\sigma$ with the results from ARIADNE.

The stellar age was derived with ARIADNE and  {\tt {PARAM1.3}} to \ageariadne~Gyr and \ageparam~Gyr, respectively. We used the stellar radius and mass from ARIADNE and   \teff~from SME in our pyaneti modelling in Sect.~\ref{Section: transit and RV modelling} and the SME abundances  for the planet interior modelling   in Sect.~\ref{subsection: interior modelling}.     
 
\section{Transit and radial velocity modelling} \label{Section: transit and RV modelling}   

\begin{figure*}[!ht]
  \centering
  \includegraphics[width=1\linewidth]{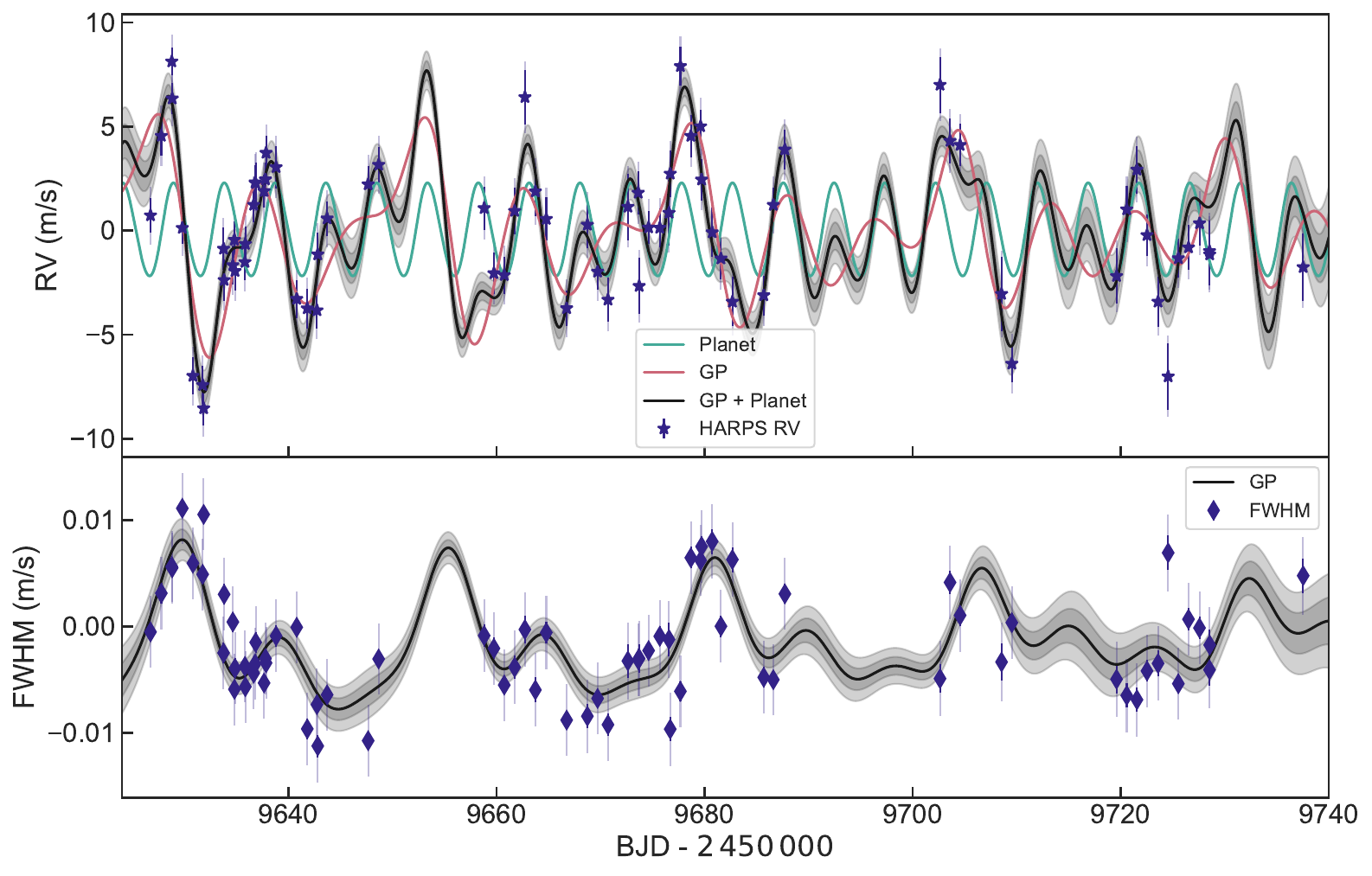} 

\caption{RV (top panel) and FWHM (bottom panel) time series. The purple markers in each panel represent the HARPS RV and FWHM measurements with inferred offsets extracted. The inferred multi-GP model is shown as a solid black curve, where the dark and light shaded areas show the 1- and 2$\sigma$ credible intervals from this model, and can also explain the data, but with a correspondingly lower probability. The solid red line in the top panel shows the star-only model, and the teal sine curve shows the Keplerian for TOI-733~b. 
In both panels, the nominal error bars are plotted in solid purple, and the jitter error bars ($\sigma_\mathrm{HARPS}$) are semi-transparent purple.}
\label{fig:RVGP}
\end{figure*}

\begin{figure}[!ht]
\centering
\resizebox{\hsize}{!}
{\includegraphics{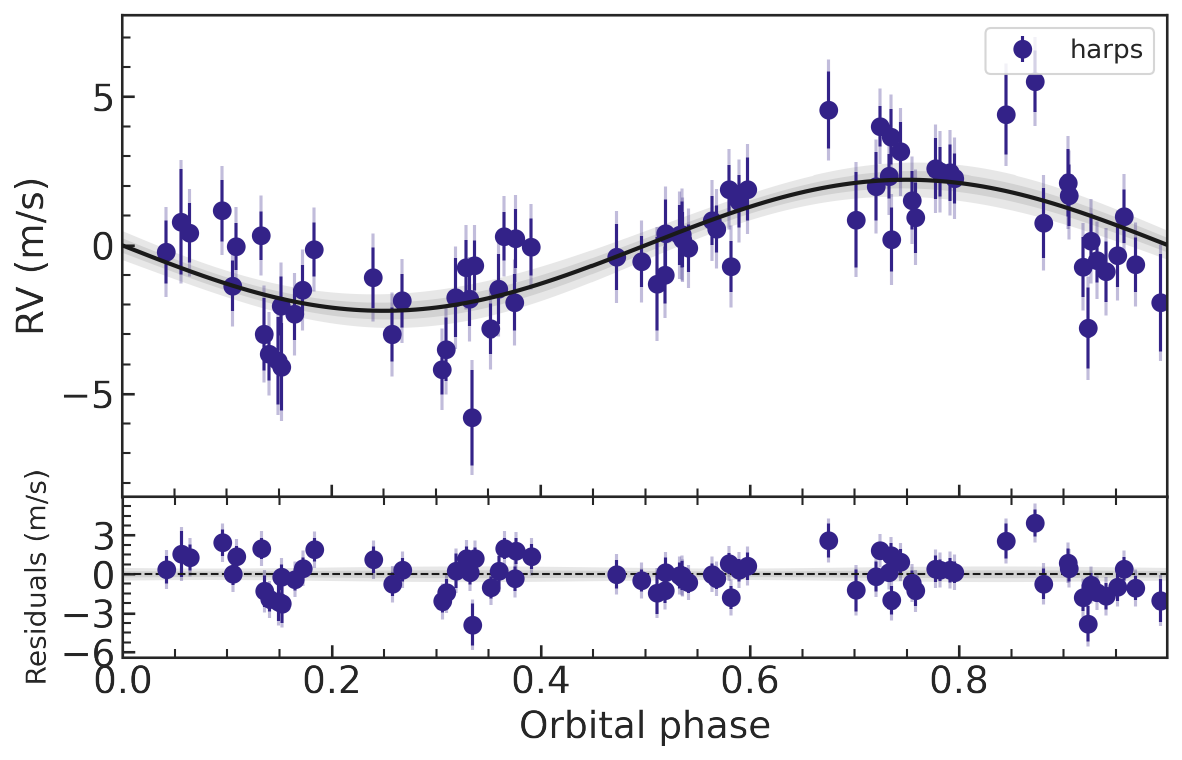}}
\caption{HARPS RV data (purple points) and inferred model (solid black curve) phase-folded on the orbital period of TOI-733~b. 1- and 2~$\sigma$ credible intervals in shaded grey regions are also shown. Nominal and jitter error bars are plotted in solid and semi-transparent purple, respectively.}
\label{Figure: RVs}
\end{figure}

 \begin{figure}[!ht]
\centering
\includegraphics[width=1\linewidth]{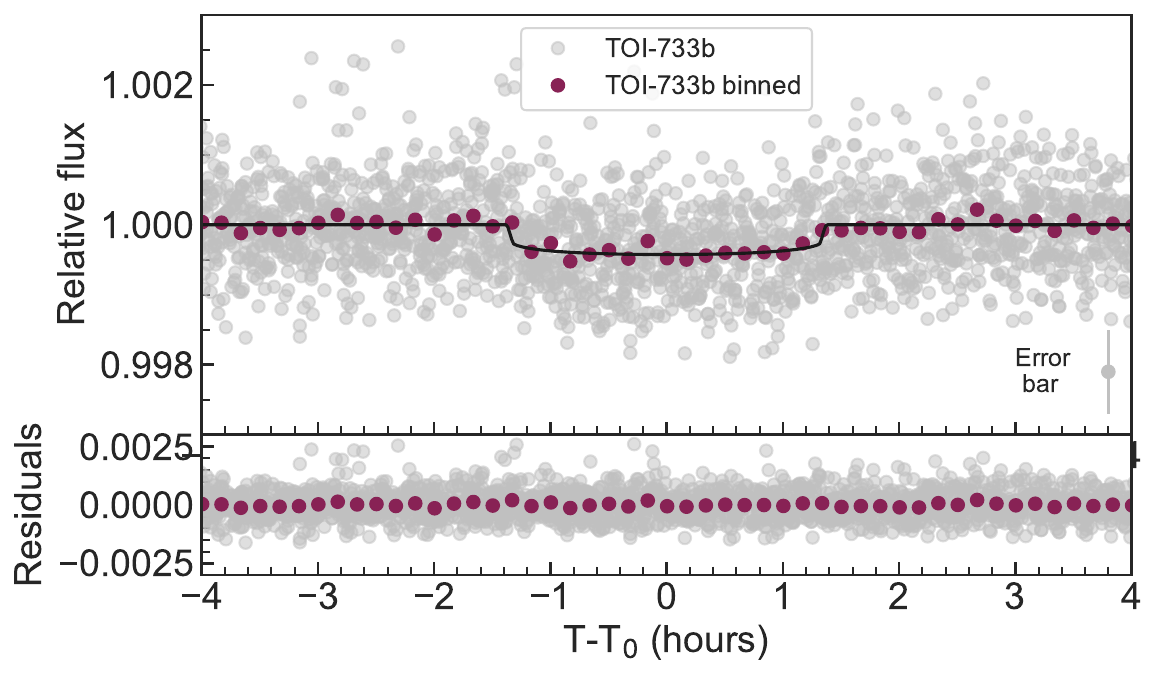}
\caption{TOI-733~b phase-folded and detrended transits from both TESS sectors, with residuals. The best-fitting transit model is marked by the black curve. Two-minute nominal cadence data points binned to 10 min are shown in grey and green, respectively. The typical error bar is added in the bottom right corner. }
\label{fig:transit}
\end{figure}

\begin{table*}[h!]
\centering
\caption{The {\tt{pyaneti}} model of TOI-733 described in Sect.~\ref{Section: transit and RV modelling}.}
\begin{tabular}{lcrr}
\hline
\hline
\noalign{\smallskip}
Parameter & Priors$\tablefootmark{a}$ & Final value  \\
\noalign{\smallskip}
\hline

\noalign{\smallskip}
             
\multicolumn{3}{l}{\emph{Fitted parameters}}\\
\noalign{\smallskip}                
~~~Transit epoch $T_0$ (BJD - 2\,450\,000)  \dotfill &  $\mathcal{U}$[8545.73, 8545.79]   & \Tzerob             \\
\noalign{\smallskip}                
~~~Orbital period \Porb \dotfill &  $\mathcal{U}$[4.8845, 4.8860]  &\Pb \\
\noalign{\smallskip}
~~~$\sqrt{e} \sin \omega_\star$  & $\mathcal{U}[-1,1]$ & \esinb[]  \\
~~~$\sqrt{e} \cos \omega_\star$ &  $\mathcal{U}[-1,1]$ & \ecosb[]  \\                
\noalign{\smallskip}                                
~~~Impact parameter $b$\dotfill &  $\mathcal{U}$[0, 1]   & \bb  \\
\noalign{\smallskip}                
~~~Scaled semi-major axis $a/R_\star$ \dotfill &  $\mathcal{N}$[14, 1]   &\arb   \\
 \noalign{\smallskip}                                      
~~~Scaled planet radius $R_{\mathrm{p}}/R_\star$ \dotfill & $\mathcal{U}$[0, 0.05]    &\rrb \\
\noalign{\smallskip}
~~~Doppler semi-amplitude variation $K $ \dotfill &   $\mathcal{U}$[0, 50]     &\kb     \\
 
\noalign{\smallskip}  

~~~Limb-darkening coefficient $q_1$ \dotfill &  $\mathcal{U}$[0, 1]   & \qone    \\              
\noalign{\smallskip}                              
~~~Limb-darkening coefficient $q_2$ \dotfill &  $\mathcal{U}$[0, 1]   & \qtwo        \\

\noalign{\smallskip}
\hline
\noalign{\smallskip}
\multicolumn{3}{l}{\emph{GP hyperparameters}}\\   
\noalign{\smallskip}

~~~GP Period $P_{\rm GP}$ \dotfill  &  $\mathcal{U}[24.5,26.5]$ & \PGP \\
~~~$\lambda_{\rm p}$  \dotfill &  $\mathcal{U}[0.1,3]$ &  \lambdap \\
~~~$\lambda_{\rm e}$  \dotfill &  $\mathcal{U}[1,200]$ &  \lambdae \\
~~~$V_{\rm c}$   \dotfill &  $\mathcal{U}[0,100]$ & \jArvc \\
~~~$V_{\rm r}$  \dotfill &  $\mathcal{U}[0,500]$ & \jArvr \\
~~~$F_{\rm c}$   \dotfill &  $\mathcal{U}[0,150]$ & \jAfwhm \\

\noalign{\smallskip}
\hline
\noalign{\smallskip}
\multicolumn{3}{l}{\emph{Derived Parameters}}\\               
\noalign{\smallskip}                                             
~~~Planet mass \mplanet\ \dotfill   & \dots     & \mpb  \\
\noalign{\smallskip}                
~~~Planet radius \rplanet\ \dotfill &  \dots    & \rpb \\
\noalign{\smallskip}
~~~Inclination $i$ \dotfill &   \dots   & \ib  \\ 
\noalign{\smallskip}                
~~~Eccentricity $e$  \dotfill & \dots & \eb  \\
\noalign{\smallskip}  
~~~Angle of periastron $\omega_\star$ \dotfill & \dots & \wb  \\              

 \noalign{\smallskip}                                              
~~~Semi-major axis $a$ \dotfill &  \dots   & \ab   \\                
 \noalign{\smallskip}                
~~~Insolation $F$ \dotfill & \dots     &  \insolationb\\
  
   \noalign{\smallskip}                
~~~Planet density \rhoplanet\ \dotfill &  \dots   & \denpb \\

 \noalign{\smallskip}                
~~~Planet  surface gravity $\log (g_\mathrm{b})$ \dotfill & \dots    &  \grapb \\
 
\noalign{\smallskip}                
~~~Equilibrium temperature $T_\mathrm{eq}\,\tablefootmark{b}$ \dotfill & \dots    & \Teqb \\
           
 \noalign{\smallskip}                
~~~Jeans escape parameter $\Lambda\,\tablefootmark{c} $ \dotfill & \dots    &  \jspb \\ 
           
 \noalign{\smallskip}                
~~~Transmission spectroscopy metric TSM\tablefootmark{d} \dotfill & \dots    &  \tsmb \\            
               
\noalign{\smallskip}                          
~~~Total transit duration $T_{14}$ \dotfill &  \dots   & \ttotb     \\
 
 \noalign{\smallskip}                          
~~~Full   transit  duration $T_{23}$ \dotfill & \dots    & \tfulb    \\               
  \noalign{\smallskip}                          
~~~Ingress and egress transit  duration $T_{12}$ \dotfill & \dots    &  \tegb    \\

\noalign{\smallskip}
\hline
\noalign{\smallskip}
\multicolumn{3}{l}{\emph{Additional Parameters}}\\      

\noalign{\smallskip}                              
~~~Offset RV HARPS \dotfill &  $\mathcal{U}$[ -24.3256 , -23.3089 ]    &  \harps     \\

~~~Offset FWHM \dotfill &  $\mathcal{U}$[ 6.4434 , 7.4657 ]    &  \FWHM     \\
               
\noalign{\smallskip}                              
~~~RV jitter HARPS \dotfill &$\mathcal{J}[0, 1000]$ &\jharps      \\    
\noalign{\smallskip}                              
~~~FWHM jitter \dotfill &$\mathcal{J}[0,1000]$ &\jFWHM      \\    
  \noalign{\smallskip}                              
~~~\textit{TESS} light curve jitter $\sigma_{TESS}$ ($\times 10^{-6}$) \dotfill & $\mathcal{J}[0, 1000]$ &   \jtr    \\    

\noalign{\smallskip} 

       
\noalign{\smallskip}                
\hline
\end{tabular}
\label{Table: Orbital and planetary parameters}
\tablefoot{
\tablefoottext{a}{$\mathcal{U}$[a,b] refers to uniform priors in the range  \emph{a} --  \emph{b}, and $\mathcal{J}$[a,b] to modified Jeffrey's priors \citep[Eq.~16 in][]{2005ApJ...631.1198G}.} 
\tablefoottext{b}{Dayside equilibrium temperature, assuming no heat redistribution and zero albedo.}
\tablefoottext{c}{$\Lambda =  G M_\mathrm{p} m_\mathrm{H} / (k_\mathrm{B} T_\mathrm{eq} R_\mathrm{p})$ \citep{2017A&A...598A..90F}.}
\tablefoottext{d}{\citet{2018PASP..130k4401K}.}
}
\end{table*}

\begin{figure*}[!ht]
\makebox[\textwidth][c]{\includegraphics[width=0.8\textwidth]{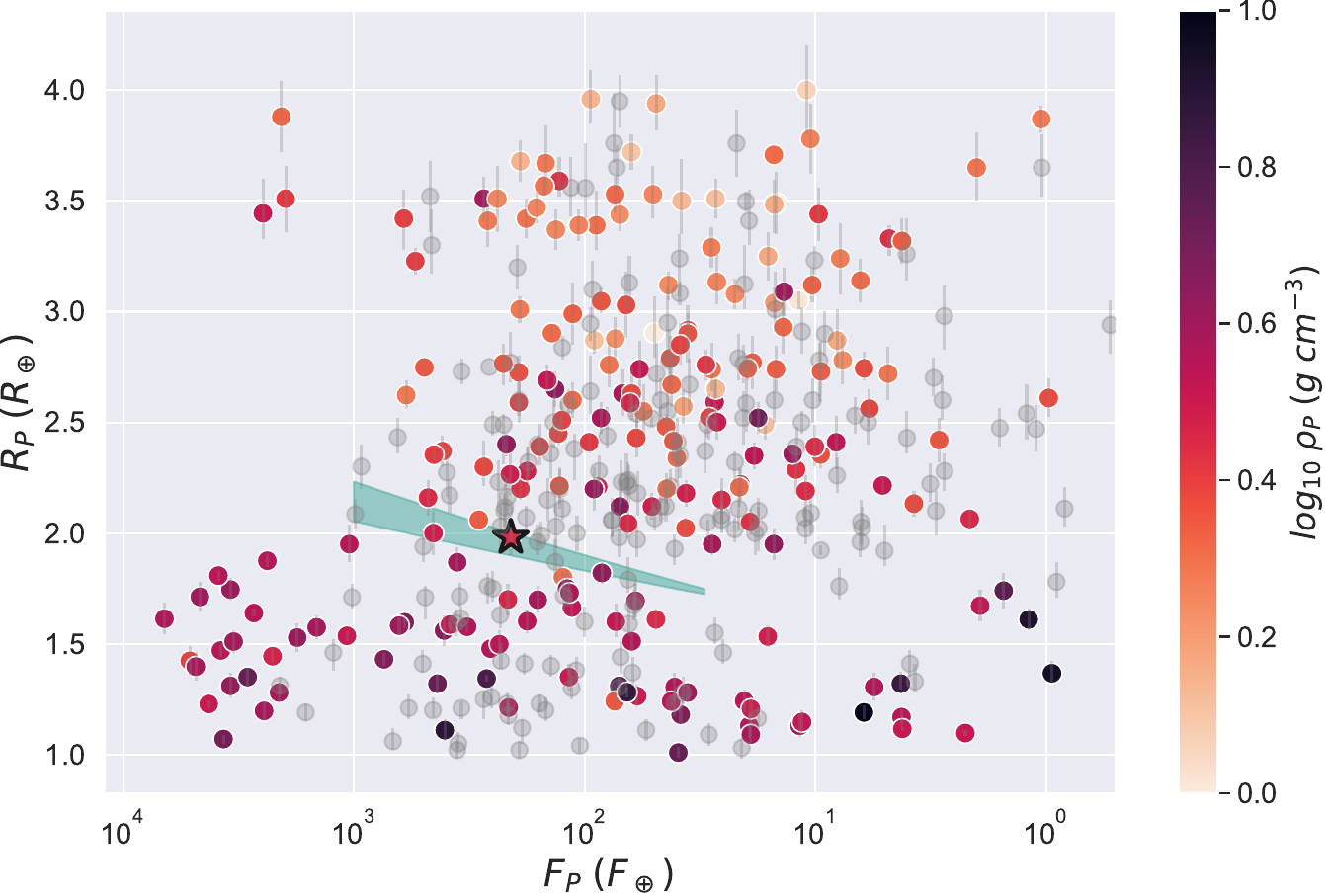}}
\caption{Radius vs incident flux  (in units of flux received on Earth) for small planets ($1-4$~\rearth) orbiting stars $0.7-1.4$~\Msun\ and radius estimates with a precision better than 5\%. All data were taken from the NASA Exoplanet 
Archive. 
Points in colour correspond to planets with density (and thus mass) estimates, where lighter and darker colours correspond to lower and higher densities, respectively. Grey points are planets whose masses have not been measured, and thus their densities are unknown. A fit to the radius valley following the relation in \citet{2022AJ....163..179P} is plotted in semi-transparent teal. TOI-733~b, marked with a black star, lies well within the sparsely populated region of the radius gap.}
\label{fig:r-f}
\end{figure*}

For the joint modelling of TOI-733, we used the code \href{https://github.com/oscaribv/pyaneti} \pyan\footnote{\url{https://github.com/oscaribv/pyaneti}}\ \citep{pyaneti, pyaneti2} to obtain and refine the system parameters. As mentioned in Sect.~\ref{Section: obs}, we only use trimmed versions of the \citla-detrended light curves from the two TESS sectors. Each segment contained 24 hours of data, including and around each transit (total transit duration $\sim$2.6~hours). We accounted for stellar limb darkening using the \citet{Kipping2013} $q_1$ and $q_2$ parametrisation, and modelled the transits using the \citet{Mandel2002} approach. The orbit inclination was estimated via the impact parameter parametrisation \citep{2010exop.book...55W}, which ultimately allowed us to estimate the true planet mass.

In contrast to what the quiet look of the light curves (Fig.~\ref{Figure: citla}) may suggest, TOI-733 has a pronounced activity signature (Sects.~ \ref{sect: freq}, \ref{Section: Stellar modelling}). We thus applied a multi-dimensional Gaussian process approach, the \pyan\ implementation of which is as described in \citet{Rajpaul2015}. The activity indicator of choice to pair with the DRS RVs and guide the GP is the FWHM because it clearly shows the imprint of the star (Sect.~\ref{sect: freq}). We tested combinations with other available activity indicators extracted via the different pipelines, but for the purpose of this analysis, none yielded superior results to the pairing with the FWHM. Because the periodocity of the star-induced signal is clear, we used the quasi-periodic (QP, Eq.~\ref{eq:gamma}) kernel and placed an uninformative prior with a range containing the value corresponding to the peak of the FWHM (and dLW) GLS periodogram ($\sim$25~days; see Fig.~\ref{fig: GLS}, fourth panel). Because the first two harmonics of this signal are clearly detected in the RV data, we consider this to be the true stellar rotation period, \Prot. We add that, while the S-index shows a significant peak suggesting a \Prot\ of $\sim$65.2~days, modelling it instead of the FWHM and adjusting the priors accordingly still converges on the same \Prot\ as was given by the FWHM. This further confirms our conclusion.  

The $P_{\rm GP}$ term in Eq.~\ref{eq:gamma} is to be interpreted as this \Prot, while $\lambda_p$ describes (the inverse of) the harmonic complexity of the data, and $\lambda_e$ represents the time evolution of the active features as they move along the stellar surface, 

\begin{equation}
    \gamma(t_i,t_j) = \exp 
    \left[
    - \frac{\sin^2[\pi(t_i - t_j)/P_{\rm GP}]}{2 \lambda_{\rm P}^2}
    - \frac{(t_i - t_j)^2}{2\lambda_{\rm e}^2}
    \right].
    \label{eq:gamma}
\end{equation}

Similarly to  \citet{Geo21} and \citet{2022MNRAS.514.1606B}, for example, the two-dimensional GP we used to characterise the TOI-733 system is formulated as in Eq.~\ref{eq:multi} below,

\begin{equation}
    \begin{matrix}
    \Delta RV & = & V_c G(t) + V_r \dot{G}(t), \\
    \Delta FWHM & = & F_c G(t). \\
\end{matrix}
\label{eq:multi}
\end{equation}

G(t) is  assumed to describe the RV and activity indicator time series and is a latent variable modelled by the QP covariance function in Eq.~\ref{eq:gamma}. $V_c$, $V_r$ , and $F_c$ are coefficients that relate G(t) to the observables.

G(t) and $\mathrm{\dot{G}(t)}$  represent the GP function and its first derivative, respectively. The dependence of the position of the spots on the stellar hemisphere is modelled by the dG/dt part. In the case of the RVs, the latter is particularly relevant (as shown by the value of $V_r$; see Table~\ref{Table: Orbital and planetary parameters}) since RVs depend not only on the fraction of the stellar disc covered by active regions, but also on how the size and shape of these surface features change in time. 

Using the polar form parametrisation for $e$ and $\omega_\star$ and adding a jitter term for the photometric and spectroscopic data, we proceeded with the model configuration described above to sample the parameter space with 500 Markov chains. Convergence was checked at every 5000 steps, and when it was reached, the last set of 5000 was used, along with a thin factor of 10, to create posterior distributions for the sampled parameters, each built with 250 000 independent points. All parameters, the priors we used, and derived values are listed in Table~\ref{Table: Orbital and planetary parameters}. Our resulting final multi-GP model is shown in Fig.~\ref{fig:RVGP}, where the top panel shows the RV, and the bottom panel shows the FWHM time series. The phase-folded RV and transit plots of TOI-733~b are shown in Figs.~\ref{Figure: RVs} and \ref{fig:transit}, respectively. For clarity, the latter shows only 8 hours centred around the transit.

We thus find TOI-733~b to be in a circular 4.885-day orbit
around a G6 V star, which in turn has a stellar rotation period estimated as \Prot\ = 25.48 days. The activity of the star is once again shown by the value of $\lambda_p$ (\lambdap), indicating a high harmonic complexity, which in turn is a sign of rapid changes within a single rotation period. The lifetime of the active regions can also be used to infer high activity, but unfortunately,
our model is not able to constrain $\lambda_e$ well.

\section{Discussion} \label{Section: disco}

Based on the two sectors of TESS data, we obtain a planet radius
of \rplanet\ = \rpb\ (4.4 \% precision), while the HARPS RVs
yield a semi-amplitude of $K$ = \kb. These in turn give
a planet mass of \mplanet\ = \mpb\ (12 \% precision) and a bulk density of \rhoplanet\ = \denpb. With an orbital period of 4.88 days around a G6~V star, TOI-733~b is in a highly irradiated orbit ($F_\mathrm{p}$ = \insolationb[]$F_\mathrm{\oplus}$),
and, as shown in Fig.~\ref{fig:r-f}, lies in the middle of the small-planet radius valley, here calculated following the work of \citet{2022AJ....163..179P}. All planets we plotted have radii with an uncertainty in radius of 5 \% at most. The data were downloaded from the NASA Exoplanet archive, where we chose for planets with several entries the most recent results with the highest precision. In cases of similar precision, the latest publications were favoured. If stellar irradiation was not among the listed parameters, we calculated it using the following relation:

\begin{equation}
F_{\mathrm{p}}=\left(\frac{R_{\star}}{R_{\odot}}\right)^2\left(\frac{T_{e f f}}{T_{\odot}}\right)^4\left(\frac{\mathrm{AU}}{a}\right)^2 F_{\oplus}
,\end{equation}

where $F_\mathrm{p}$ is the incoming stellar flux, $T_\mathrm{eff}$ is stellar effective temperature, and $a$ is the semi-major axis. Colour-coded dots are planets with known bulk densities, while the densities of the planets in grey cannot be calculated since their masses have not yet been measured. As evident from this figure, the densities of super-Earths are higher than those of the mini-Neptune population, as the latter feature a significant volatile content.

The planet to the immediate left of TOI-733~b in Fig.~\ref{fig:r-f} and thus the closest well-characterised planet to it in this parameter space, is $\pi$ Men c \citep{2018A&A...619L..10G, 2018ApJ...868L..39H, 2022AJ....163..223H}. \citet{2020ApJ...888L..21G} reported the non-detection
of photodissociated hydrogen, suggesting that $\pi$ Men c
might instead be H$_2$O or dominated by other heavy molecules
rather than H/He. The latter hypothesis was later confirmed by further observations with the detection of, most likely escaping, \ion{Ca}{II} ions \citep{2021ApJ...907L..36G}. Despite its relatively mature age \citep[$\sim$ 4 Gyr, ][]{2020A&A...642A..31D}, atmospheric escape was expected for $\pi$ Men c as its radius (2.06 $\pm$ 0.03 \rearth) is large relative to its mass (4.52 $\pm$ 0.81 \mearth)\footnote{The radius and mass values are as taken by \citet{2020ApJ...888L..21G}. More accurate parameters have since been presented in \citet{2022AJ....163..223H}}. While still of relatively low density (\denpb\ vs 2.1 $\pm$ 0.4 \gc\ for $\pi$ Men c), it is less likely that TOI-733~b is undergoing intense atmospheric loss.

Water worlds have been put forward as a possible explanation for planets with similar parameters \citep[e.g. ][]{Zeng2019, 2021ApJ...923..247Z}. Recently, \citet{Luque22} showed that the small-planet population around M dwarfs is inconsistent with a radius gap as observed around higher-mass stars. They suggest that photoevaporation is not needed to explain the observed trends and that water worlds, forming beyond the snow line and migrating inward, are the planets that straddle the area between rocky planets and those with non-negligible envelopes. The census of well-characterised planets around Sun-like stars, however, prevented this conclusion from being extended to higher mass stars.

To try and understand TOI-733~b better and elucidate its composition, and whether it is more likely that its atmosphere has or is in a process of being lost, or if instead it formed more or less as we currently find it, we performed interior and atmospheric modelling. We describe this in the following sections.

\subsection{Interior structure}

\label{subsection: interior modelling}

  \begin{figure*}[!ht]
  \makebox[\textwidth][c]{\includegraphics[width=0.8\textwidth]
  {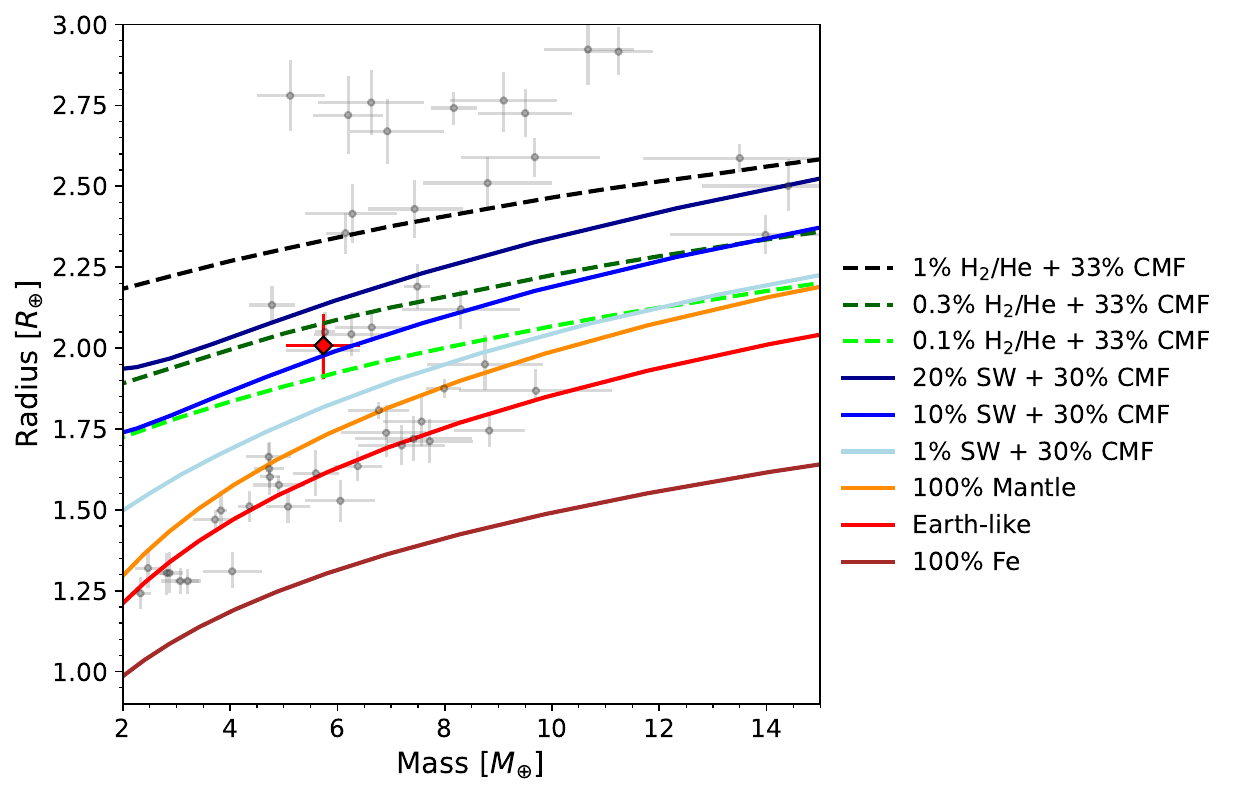}}
   \caption{Mass-radius relations for SW planets \citep{Acuna21, Agui2021}, planets with Earth-like cores and H/He envelopes \citep{LopFor14}, and rocky planets with different iron contents \citep[bottom three curves;][]{Brugger17}. The atmospheres in volatile-rich planets are in radiative equilibrium for irradiation temperatures of 1200 K and 1000 K for water and H/He envelopes, respectively. Assuming an age of 4.4~Gyr, the position of TOI-733~b is as highlighted in red. Grey points correspond to planets less massive than 15~$M_{\oplus}$ with mass and radius data available from the NASA Exoplanet Archive. All planets have a limit on the radius uncertainty of 5~\%, and a limit on the mass of 15~\%. The mantle composition follows the model of \citet{Brugger16} with both CMF and WMF equal to zero, and is made up of silicate rock. Earth has a CMF of 0.32 and a WMF of 0.0005.}

      \label{Figure: MR_diagram}
 \end{figure*}

To illustrate the position of TOI-733 b in mass-radius space, we show in Fig.~\ref{Figure: MR_diagram} the iso-composition curves for refractory interiors \citep{Brugger16, Brugger17}, planets with supercritical water (SW) envelopes \citep{Acuna21, Agui2021}, and planets with H/He envelopes \citep{LopFor14}. We choose to use the data grid of \citet{LopFor14} to plot different percentages of H/He models over the more widely used model of \citet{Zeng2019} because the latter indicate that the temperature in \citet{Zeng2019} is that at the $P=100$ bar level, whereas the temperature in the model of \citet{LopFor14} reflects the irradiation or equilibrium temperature of the planet. This concept is further elaborated in \citet{Rogers2023-ApJL}, for example.

 Fig.~\ref{Figure: MR_diagram} shows that the density of TOI-733~b is lower than that of a pure mantle rock planet, suggesting that it contains a volatile layer. A 5.7~$M_{\oplus}$ planet with a H/He-dominated envelope of  $\sim$0.2~\%  has a radius of $\simeq  2~R_{\oplus}$ \citep{LopFor14}. Therefore, with a radius of $R =  2.0 \ R_{\oplus}$, the most likely inventory of volatiles in TOI-733~b does not include a significant H/He component and is instead that of a secondary atmosphere \citep[H$_{2}$O, CO$_{2}$, CH$_4$, etc.; ][]{Madhu21, Krissansen-Totton_2022}, which is the envelope composition we assumed in our interior structure model.

We performed a Markov chain Monte Carlo (MCMC) Bayesian analysis \citep{Acuna_submitted,Director17} of the interior structure and composition of TOI-733~b. Our 1D interior structure model considered three layers: a Fe-rich core, a silicate mantle \citep{Brugger16,Brugger17}, and a water-dominated envelope in supercritical and steam phases, given the high irradiation TOI-733~b receives from its host star \citep{Mousis20,Acuna21}. To include the effect of this high irradiation on the total radius self-consistently, we coupled our interior model to an atmospheric model that computed the emitted total radiation and reflection of the atmosphere to determine radiative-convective equilibrium \citep{Acuna21,Acuna_submitted}. Our interior-atmosphere models calculated the radius from the centre of the planet up to a transit radius of 20 mbar \citep{Grimm18,Mousis20}. 

In our analysis, we considered two scenarios. Scenario 1 is the most conservative because it only takes the planetary mass and radius into account as input for the MCMC method, whereas in scenario 2, we adopted as input for the MCMC the stellar Fe/Si mole ratio in addition to the mass and radius of the planet. We obtain a Fe/Si = 0.67 $\pm$ 0.11, following the approach described in \citet{Brugger17,Sotin07} to convert the stellar abundances in Table \ref{Table: Star basic parameters} into a mole ratio. We adopted solar composition reference values from \cite{Gray_book05}. The MCMC provides the posterior distribution functions (PDF) of the compositional parameters, which are the core mass fraction (CMF) and the water mass fraction (WMF). In addition, the atmospheric parameters were also obtained by the MCMC and consist of the temperature at the interior-atmosphere coupling interface (300 bar), the Bond albedo, and the atmospheric thickness from 300 bar up to the transit radius. 

\begin{table}[h]
\centering
\caption{1$\sigma$ confidence intervals of the interior and atmosphere MCMC output parameters in the two different compositional scenarios (see text).}
\begin{tabular}{lcc}
\hline
Parameter & Scenario 1 & Scenario 2 \\ \hline
Core mass fraction, CMF & 0.27$\pm$0.14 & 0.20$\pm$0.03 \\
Water mass fraction, WMF & 0.11$\pm$0.06 & 0.07$\pm$0.05 \\
Fe-to-Si mole ratio, Fe/Si & 1.11$\pm$0.74 & 0.67$\pm$0.11 \\
Temperature at 300 bar, T$_{300}$ [K] & 3458$\pm$38 & 3448$\pm$30 \\
Thickness at 300 bar, z$_{300}$ [km] & 616 $^{+92}_{-185}$ & 447 $^{+111}_{-21}$ \\
Albedo, a$_{p}$& \multicolumn{2}{c}{0.24$\pm$0.01} \\
Core+Mantle radius, [R$_{p}$ units] & 0.77$\pm$0.06 & 0.86$^{+0.07}_{-0.02}$ \\ \hline
\end{tabular}
\label{tab:mcmc_interior}
\end{table}

Table \ref{tab:mcmc_interior} shows the mean and 1$\sigma$ confidence intervals of the MCMC output parameters. In scenario 1, which is the most general and conservative scenario because we do not make any assumptions on the planetary Fe/Si mole ratio, the CMF distribution is centred at a similar value to the mean of the CMF distribution of the rocky super-Earth population \citep{Plotnykov20}. In addition, in scenario 1 the CMF is compatible within the uncertainties with the Earth CMF value (CMF$_{\oplus} = 0.32$). The CMF in scenario 2 is significantly lower than that of Earth, which is a consequence of a lower Fe/Si mole ratio of the stellar host compared to the Sun (Fe/Si$_{\odot}$ = 0.96), although the planetary CMF is still well within the range of CMFs observed in super-Earths ($\simeq$ 0.10 to 0.50). The WMF of TOI-733~b ranges from 5 to 17\% in scenario 1 and from 2 to 12\% in scenario 2, suggesting that the water content of TOI-733~b is in between to what is expected in super-Earths (WMF < 5\%) and sub-Neptunes (WMF > 20\%) \citep{Acuna_inprep,Luque22}.

\subsection{Atmospheric escape}

The low surface gravity of TOI-733~b combined with its high equilibrium temperature results in a moderately low value of the Jeans escape parameter $\Lambda=20.6$. Neptune-like planets for which $\Lambda\lesssim 20$ are expected to have quickly escaping atmospheres \citep{2016ApJ...817..107O,Cubillos2017}. Their size would decrease to smaller radii, so that $\Lambda$ would increase to reach higher values, reducing atmospheric escape rates \citep{2017A&A...598A..90F}.

To quantify this effect, we followed the approach from \cite{Agui2021} to estimate the total mass of H/He that TOI-733~b may have had in the past. The photoevaporation mass-loss rate from the atmosphere in the energy-limited regime is \citep{Er07,2013ApJ...775..105O}

\begin{equation}
    \dot{M}=\epsilon \frac{\pi F_{\mathrm{XUV}} R_{\mathrm{p}}^3}{G M_{\mathrm{p}}},
\end{equation}

where $F_\mathrm{XUV}$ is the XUV flux received by the planet, $G$ is the gravitational constant, and $\epsilon$ is an efficiency parameter. We approximated the XUV luminosity by the analytical fit obtained by \cite{Sanz2011}, and we estimate $\epsilon\simeq 0.07$ from \cite{Owen2012}. Since the XUV luminosity is a decreasing function of time, the mass-loss rate also decreases with time. This yields a mass loss of $9.3 \times 10^{-1}$ \mearth/Gyr during the saturation regime and a present-day mass-loss rate of $2.1 \times 10^{-3}$ \mearth/Gyr at the estimated age of 4.4 Gyr. Following the approach of \cite{Agui2021}, we estimate the total mass of H/He lost by photoevaporation by integrating the mass-loss rate, assuming that $M_\mathrm{p}$, $R_\mathrm{p}$ and $T_\mathrm{eq}$ remained roughly constant, and that only the XUV flux decreased during the planetary evolution. In this case, we find that TOI-733~b could have lost $\sim 0.12$ \mearth\ of H/He, that is, $\sim 2\%$ of its initial mass. This estimate is consistent with the computation made by \cite{Rogers2023-ApJL}, who predicted that at $T_\mathrm{eq}=800$ K, planets with core masses $\lesssim 6$~\mearth\ are entirely stripped of their envelopes, assuming the latter are made of pure \ce{H2}. This is expected to remain valid at higher equilibrium temperatures.

Despite the efficient hydrogen escape at early ages, recent studies showed that a secondary atmosphere may be formed by outgassing volatile gases from the magma after the photoevaporation phase \citep{KiteandBarnett2020pnas,TianandHeng2023}. The present-day mass-loss rate by photoevaporation may be insufficient to remove the outgassed hydrogen due to the low XUV flux from the host star. Nevertheless, other mechanisms of thermal escape might cause the preferential loss of hydrogen. The Jeans parameter of TOI-733~b is lower than that of Earth ($\Lambda_\oplus = 27.4$), which results in a hydrogen Jeans escape rate $\sim 10^3$ times greater for TOI-733~b than for Earth \citep[see][for the Jeans escape rate formula]{CatlingandKasting2017book}. It is therefore very likely that any outgassed hydrogen was removed by thermally driven escape, leaving behind a secondary atmosphere made of heavier volatiles as on Earth.

This supports the hypothesis that TOI-733~b may have formed with an envelope that is a mixture of various volatile compounds, but only heavier species remained after the escape of H/He. In other words, the planet may have formed with H/He and water, but H/He was lost, and what is currently left behind is a mixture of the initial water reservoir together with any outgassed gases.

Furthermore, it is possible that TOI-733~b formed with more than $2\%$ of H/He by mass. As shown in Figure \ref{Figure: MR_diagram}, H/He envelopes are very inflated at these high temperatures, meaning that using the present-day radius underestimates the mass-loss rate. 

A further possibility is that TOI-733~b directly formed as an ocean planet and did not experience atmospheric loss because water has a much lower escape efficiency \citep{2021MNRAS.502..750I}. In other words, the planet formed with an initial high WMF and was able to retain it because water is more resistant to XUV photoevaporation than H/He.

In both cases, the loss of the entire H/He content from the atmosphere of TOI-733~b supports the presence of a secondary atmosphere that might be dominated by water. However, atmospheres of other heavy volatiles \citep{2015ApJ...807....8H,2017MNRAS.464.3728B,2021MNRAS.502..750I} cannot be excluded.

\subsection{Prospects for atmospheric characterisation}

Based on our interior structure analysis, TOI-733~b most probably features a volatile envelope. The composition of this envelope is likely to be that of a secondary atmosphere, although the presence of a few tens of percent of H or He cannot be completely ruled out. Therefore, TOI-733~b is an interesting target for atmosphere characterisation observations to confirm that its atmosphere is dominated by H$_{2}$O, CO$_{2}$, CH$_{4}$ or other compounds that are present in a secondary atmosphere instead of H/He. These observations would enable us to break the degeneracy between envelope mass and composition that is typically found in sub-Neptunes. Unfortunately, the estimated TSM (transmission spectroscopy metric) and ESM (emission spectroscopy metric) of TOI-733~b are 46.29 and 4.4, respectively, which both place it below the threshold of the optimal targets for transmission and emission spectroscopy with the James Webb Space Telescope \citep{Kempton18}. However, an extended atmosphere signature could be significantly larger than the TSM would imply because the latter is based on the assumption of a bound atmosphere. It is thus worth mentioning that a search from the ground for an H$\alpha$ or \ion{He}{I} signature, of any extended, escaping H/He atmosphere, or possibly even H from the photodissociated ocean world, might be possible \citep[e.g. ][]{Jensen12, 2017AJ....153..217C}. 
Any attempts to observe this planet in the hope of learning more about its bound atmosphere, however, will have to be postponed until the next generation of telescopes.

\section{Conclusions} \label{Section: tada}  

We presented the discovery and characterisation of TOI-733~b. Our stellar and joint RV and transit modelling shows that this planet orbits a G6~V star and is located well within the small-planet radius valley when considering solar-type stars. We performed interior and atmospheric modelling to try and narrow down the possible structure and composition of this planet. We found that if TOI-733~b ever had a H/He atmosphere, it was mostly if not completely lost, leaving behind a secondary atmosphere of heavier elements. Our analysis also indicates that the planet may also have formed as a water world and did not experience atmospheric mass loss.

Answering the question of whether TOI-733~b has a secondary atmosphere or is an ocean planet means that we need to distinguish between a Neptune-like planet that lost $\sim$10~\% of its H/He (as estimated by our atmospheric mass loss models) to leave behind a steam atmosphere of heavier volatiles, and a planet that formed and remained relatively the same throughout its evolution. While being beyond the scope of this paper, finding an answer to this question will have broad implications for our understanding of exoplanets.

The similarity between TOI-733~b and $\pi$~Men~c does not end at the connection of the radius to the incident flux. In addition to receiving a similar amount of stellar irradiation, the two planets orbit stars of similar type and age. Models suggest that H$_2$O plays a significant role in the interior and possibly in the envelope of both planets. Recent transmission spectroscopy observations point to an increasing probability that this is the case. Given the observability limitations of TOI-733~b, a more detailed comparison study between the two planets and their hosts could help determine to what extent, if at all, the conclusions derived for $\pi$~Men~c can be extended to TOI-733~b. If these two and other planets with similar characteristics are confirmed to indeed be dominated by water, this could point to there being a population of planets that belongs in the radius gap and are not just passing through. This does not diminish the importance of or the need for mechanisms that explain atmospheric loss, but it may mean that they and the planets considered to be or to have been subjected to them at some point in their history need to be rethought. Regardless of the case of TOI-733~b, however, because both the core-powered mass loss (formation) and the XUV photoevaporation (evolution) mechanisms are able to physically explain the presence of the radius valley separating super-Earths from mini-Neptunes, well-characterised planets in this parameter space are essential to facilitate understanding of which of these mechanisms is dominant.

By all accounts, TOI-733~b is an interesting planet and holds the potential of being a small but key piece to solving major puzzles in exoplanet science. With ever-increasing detailed theoretical analyses and the promise of high-precision follow-up by current and upcoming facilities, we seem to be well on the way to finding answers to the main questions relating to planet formation and evolution.

\begin{acknowledgements}

This paper includes data collected by the TESS mission. Funding for the TESS mission is provided by the NASA Explorer Program. We acknowledge the use of public TOI Release data from pipelines at the TESS Science Office and at the TESS Science Processing Operations Center. Resources supporting this work were provided by the NASA High-End Computing (HEC) Program through the NASA Advanced Supercomputing (NAS) Division at Ames Research Center for the production of the SPOC data products. This research has made use of the Exoplanet Follow-up Observation Program website, which is operated by the California Institute of Technology, under contract with the National Aeronautics and Space Administration under the Exoplanet Exploration Program.

This work has made use of data from the European Space Agency (ESA) mission {\it Gaia} (\url{https://www.cosmos.esa.int/gaia}), processed by the {\it Gaia} Data Processing and Analysis Consortium (DPAC, \url{https://www.cosmos.esa.int/web/gaia/dpac/consortium}). Funding for the DPAC has been provided by national institutions, in particular the institutions participating in the {\it Gaia} Multilateral Agreement.

This work was supported by the KESPRINT collaboration, an international consortium devoted to the characterization and research of exoplanets discovered with space-based missions (\url{http://www.kesprint.science}). 

Data availability: The data underlying this article are available in machine-readable format at the CDS, as well as ExoFOP-TESS\footnote{\url{https://exofop.ipac.caltech.edu/tess}}.

I.Y.G., C.M.P., J.K, and M.F. gratefully acknowledge the support of the Swedish National Space Agency (DNR 174/18, 65/19, 2020-00104, 177/19).

K.W.F.L. was supported by Deutsche Forschungsgemeinschaft grants RA714/14-1 within the DFG Schwerpunkt SPP 1992, Exploring the Diversity of Extrasolar Planets.

This material is based upon work supported by NASA’S Interdisciplinary Consortia for Astrobiology Research (NNH19ZDA001N-ICAR) under award number 19-ICAR19\_2-0041.

K.A.C. acknowledges support from the TESS mission via subaward s3449 from MIT.
This work makes use of observations from the LCOGT network. Part of the LCOGT telescope time was granted by NOIRLab through the Mid-Scale Innovations Program (MSIP). MSIP is funded by NSF.\\

Some of the observations in this paper made use of the High-Resolution Imaging instrument Zorro and were obtained under Gemini LLP Proposal Number: GN/S-2021A-LP-105. Zorro was funded by the NASA Exoplanet Exploration Program and built at the NASA Ames Research Center by Steve B. Howell, Nic Scott, Elliott P. Horch, and Emmett Quigley. Zorro was mounted on the Gemini South telescope of the international Gemini Observatory, a program of NSF’s OIR Lab, which is managed by the Association of Universities for Research in Astronomy (AURA) under a cooperative agreement with the National Science Foundation. on behalf of the Gemini partnership: the National Science Foundation (United States), National Research Council (Canada), Agencia Nacional de Investigación y Desarrollo (Chile), Ministerio de Ciencia, Tecnología e Innovación (Argentina), Ministério da Ciência, Tecnologia, Inovações e Comunicações (Brazil), and Korea Astronomy and Space Science Institute (Republic of Korea).\\

R.L. acknowledges funding from University of La Laguna through the Margarita Salas Fellowship from the Spanish Ministry of Universities ref. UNI/551/2021-May 26, and under the EU Next Generation funds.

HD acknowledges support from the Spanish Research Agency of the Ministry of Science and Innovation (AEI-MICINN) under the grant with reference PID2019-107061GB-C66, DOI: 10.13039/501100011033.

This paper includes data collected by the TESS mission. Funding for the TESS mission is provided by the NASA Explorer Program. We acknowledge the use of public TOI Release data from pipelines at the TESS Science Office and at the TESS Science Processing Operations Center. Resources supporting this work were provided by the NASA High-End Computing (HEC) Program through the NASA Advanced Supercomputing (NAS) Division at Ames Research Center for the production of the SPOC data products. 
This research has made use of the Exoplanet Follow-up Observation Program (ExoFOP; DOI: 10.26134/ExoFOP5) website, which is operated by the California Institute of Technology, under contract with the National Aeronautics and Space Administration under the Exoplanet Exploration Program.\\

This work has made use of data from the European Space Agency (ESA) mission {\it Gaia} (\url{https://www.cosmos.esa.int/gaia}), processed by the {\it Gaia} Data Processing and Analysis Consortium (DPAC, \url{https://www.cosmos.esa.int/web/gaia/dpac/consortium}). Funding for the DPAC has been provided by national institutions, in particular the institutions participating in the {\it Gaia} Multilateral Agreement.\\

This research has made use of the NASA Exoplanet Archive, which is operated by the California Institute of Technology, under contract with the National Aeronautics and Space Administration under the Exoplanet Exploration Program.

\end{acknowledgements}

\bibliographystyle{aa}
\bibliography{references}

\begin{thebibliography}{119}
\expandafter\ifx\csname natexlab\endcsname\relax\def\natexlab#1{#1}\fi

\bibitem[{{Acu\~na} {et~al.}(2021){Acu\~na}, {Deleuil, Magali}, {Mousis,
  Olivier}, {Marcq, Emmanuel}, {Levesque, Ma\"eva}, \& {Aguichine,
  Artyom}}]{Acuna21}
{Acu\~na}, L., {Deleuil, Magali}, {Mousis, Olivier}, {et~al.} 2021, A\&A, 647,
  A53

\bibitem[{Acuña(in prep.{\natexlab{a}})}]{Acuna_submitted}
Acuña, L. in prep.{\natexlab{a}}

\bibitem[{Acuña(in prep.{\natexlab{b}})}]{Acuna_inprep}
Acuña, L. in prep.{\natexlab{b}}

\bibitem[{{Aguichine} {et~al.}(2021){Aguichine}, {Mousis}, {Deleuil}, \&
  {Marcq}}]{Agui2021}
{Aguichine}, A., {Mousis}, O., {Deleuil}, M., \& {Marcq}, E. 2021, \apj, 914,
  84

\bibitem[{{Allard} {et~al.}(2012){Allard}, {Homeier}, \&
  {Freytag}}]{2012RSPTA.370.2765A}
{Allard}, F., {Homeier}, D., \& {Freytag}, B. 2012, Philosophical Transactions
  of the Royal Society of London Series A, 370, 2765

\bibitem[{{Ambikasaran} {et~al.}(2016){Ambikasaran}, {Foreman-Mackey},
  {Greengard}, {Hogg}, \& {O’Neil}}]{Ambi2016}
{Ambikasaran}, S., {Foreman-Mackey}, D., {Greengard}, L., {Hogg}, D.~W., \&
  {O’Neil}, M. 2016, IEEE Transactions on Pattern Analysis and Machine
  Intelligence, 38, 252

\bibitem[{{Anglada-Escud{\'e}} \& {Butler}(2012)}]{2012ApJS..200...15A}
{Anglada-Escud{\'e}}, G. \& {Butler}, R.~P. 2012, \apjs, 200, 15

\bibitem[{{Baglin} {et~al.}(2006){Baglin}, {Auvergne}, {Boisnard}, {Lam-Trong},
  {Barge}, {Catala}, {Deleuil}, {Michel}, \& {Weiss}}]{Baglin2006}
{Baglin}, A., {Auvergne}, M., {Boisnard}, L., {et~al.} 2006, in 36th COSPAR
  Scientific Assembly, Vol.~36, 3749

\bibitem[{{Barrag{\'a}n} {et~al.}(2021){Barrag{\'a}n}, {Aigrain}, {Gillen}, \&
  {Guti{\'e}rrez-Canales}}]{Oscar-letter}
{Barrag{\'a}n}, O., {Aigrain}, S., {Gillen}, E., \& {Guti{\'e}rrez-Canales}, F.
  2021, Research Notes of the American Astronomical Society, 5, 51

\bibitem[{{Barrag{\'a}n} {et~al.}(2022{\natexlab{a}}){Barrag{\'a}n}, {Aigrain},
  {Rajpaul}, \& {Zicher}}]{pyaneti2}
{Barrag{\'a}n}, O., {Aigrain}, S., {Rajpaul}, V.~M., \& {Zicher}, N.
  2022{\natexlab{a}}, \mnras, 509, 866

\bibitem[{{Barrag{\'a}n} {et~al.}(2022{\natexlab{b}}){Barrag{\'a}n},
  {Armstrong}, {Gandolfi}, {Carleo}, {Vidotto}, {Villarreal D'Angelo},
  {Oklop{\v{c}}i{\'c}}, {Isaacson}, {Oddo}, {Collins}, {Fridlund}, {Sousa},
  {Persson}, {Hellier}, {Howell}, {Howard}, {Redfield}, {Eisner}, {Georgieva},
  {Dragomir}, {Bayliss}, {Nielsen}, {Klein}, {Aigrain}, {Zhang}, {Teske},
  {Twicken}, {Jenkins}, {Esposito}, {Van Eylen}, {Rodler}, {Adibekyan},
  {Alarcon}, {Anderson}, {Akana Murphy}, {Barrado}, {Barros}, {Benneke},
  {Bouchy}, {Bryant}, {Butler}, {Burt}, {Cabrera}, {Casewell}, {Chaturvedi},
  {Cloutier}, {Cochran}, {Crane}, {Crossfield}, {Crouzet}, {Collins}, {Dai},
  {Deeg}, {Deline}, {Demangeon}, {Dumusque}, {Figueira}, {Furlan}, {Gnilka},
  {Goad}, {Goffo}, {Guti{\'e}rrez-Canales}, {Hadjigeorghiou}, {Hartman},
  {Hatzes}, {Harris}, {Henderson}, {Hirano}, {Hojjatpanah}, {Hoyer},
  {Kab{\'a}th}, {Korth}, {Lillo-Box}, {Luque}, {Marmier}, {Mo{\v{c}}nik},
  {Muresan}, {Murgas}, {Nagel}, {Osborne}, {Osborn}, {Osborn}, {Palle},
  {Raimbault}, {Ricker}, {Rubenzahl}, {Stockdale}, {Santos}, {Scott},
  {Schwarz}, {Shectman}, {Raimbault}, {Seager}, {S{\'e}gransan}, {Serrano},
  {Skarka}, {Smith}, {{\v{S}}ubjak}, {Tan}, {Udry}, {Watson}, {Wheatley},
  {West}, {Winn}, {Wang}, {Wolfgang}, \& {Ziegler}}]{2022MNRAS.514.1606B}
{Barrag{\'a}n}, O., {Armstrong}, D.~J., {Gandolfi}, D., {et~al.}
  2022{\natexlab{b}}, \mnras, 514, 1606

\bibitem[{{Barrag{\'a}n} {et~al.}(2019){Barrag{\'a}n}, {Gandolfi}, \&
  {Antoniciello}}]{pyaneti}
{Barrag{\'a}n}, O., {Gandolfi}, D., \& {Antoniciello}, G. 2019, \mnras, 482,
  1017

\bibitem[{{Bolmont} {et~al.}(2017){Bolmont}, {Selsis}, {Owen}, {Ribas},
  {Raymond}, {Leconte}, \& {Gillon}}]{2017MNRAS.464.3728B}
{Bolmont}, E., {Selsis}, F., {Owen}, J.~E., {et~al.} 2017, \mnras, 464, 3728

\bibitem[{{Borucki} {et~al.}(2010){Borucki}, {Koch}, {Basri}, {Batalha},
  {Brown}, {Caldwell}, {Caldwell}, {Christensen-Dalsgaard}, {Cochran},
  {DeVore}, {Dunham}, {Dupree}, {Gautier}, {Geary}, {Gilliland}, {Gould},
  {Howell}, {Jenkins}, {Kondo}, {Latham}, {Marcy}, {Meibom}, {Kjeldsen},
  {Lissauer}, {Monet}, {Morrison}, {Sasselov}, {Tarter}, {Boss}, {Brownlee},
  {Owen}, {Buzasi}, {Charbonneau}, {Doyle}, {Fortney}, {Ford}, {Holman},
  {Seager}, {Steffen}, {Welsh}, {Rowe}, {Anderson}, {Buchhave}, {Ciardi},
  {Walkowicz}, {Sherry}, {Horch}, {Isaacson}, {Everett}, {Fischer}, {Torres},
  {Johnson}, {Endl}, {MacQueen}, {Bryson}, {Dotson}, {Haas}, {Kolodziejczak},
  {Van Cleve}, {Chandrasekaran}, {Twicken}, {Quintana}, {Clarke}, {Allen},
  {Li}, {Wu}, {Tenenbaum}, {Verner}, {Bruhweiler}, {Barnes}, \&
  {Prsa}}]{2010Sci...327..977B}
{Borucki}, W.~J., {Koch}, D., {Basri}, G., {et~al.} 2010, Science, 327, 977

\bibitem[{{Brown} {et~al.}(2013){Brown}, {Baliber}, {Bianco}, {Bowman},
  {Burleson}, {Conway}, {Crellin}, {Depagne}, {De Vera}, {Dilday}, {Dragomir},
  {Dubberley}, {Eastman}, {Elphick}, {Falarski}, {Foale}, {Ford}, {Fulton},
  {Garza}, {Gomez}, {Graham}, {Greene}, {Haldeman}, {Hawkins}, {Haworth},
  {Haynes}, {Hidas}, {Hjelstrom}, {Howell}, {Hygelund}, {Lister}, {Lobdill},
  {Martinez}, {Mullins}, {Norbury}, {Parrent}, {Paulson}, {Petry}, {Pickles},
  {Posner}, {Rosing}, {Ross}, {Sand}, {Saunders}, {Shobbrook}, {Shporer},
  {Street}, {Thomas}, {Tsapras}, {Tufts}, {Valenti}, {Vander Horst}, {Walker},
  {White}, \& {Willis}}]{Brown:2013}
{Brown}, T.~M., {Baliber}, N., {Bianco}, F.~B., {et~al.} 2013, Publications of
  the Astronomical Society of the Pacific, 125, 1031

\bibitem[{{Brugger} {et~al.}(2017){Brugger}, {Mousis}, {Deleuil}, \&
  {Deschamps}}]{Brugger17}
{Brugger}, B., {Mousis}, O., {Deleuil}, M., \& {Deschamps}, F. 2017, The
  Astrophysical Journal, 850, 93

\bibitem[{Brugger {et~al.}(2016)Brugger, Mousis, Deleuil, \&
  Lunine}]{Brugger16}
Brugger, B., Mousis, O., Deleuil, M., \& Lunine, J.~I. 2016, The Astrophysical
  Journal, 831, L16

\bibitem[{{Bruntt} {et~al.}(2008){Bruntt}, {De Cat}, \& {Aerts}}]{bruntt08}
{Bruntt}, H., {De Cat}, P., \& {Aerts}, C. 2008, \aap, 478, 487

\bibitem[{{Castelli} \& {Kurucz}(2004)}]{Castelli2004}
{Castelli}, F. \& {Kurucz}, R.~L. 2004, astro-ph/0405087
  [\eprint{astro-ph/0405087}]

\bibitem[{{Catling} \& {Kasting}(2017)}]{CatlingandKasting2017book}
{Catling}, D.~C. \& {Kasting}, J.~F. 2017, {Atmospheric Evolution on Inhabited
  and Lifeless Worlds}

\bibitem[{{Cauley} {et~al.}(2017){Cauley}, {Redfield}, \&
  {Jensen}}]{2017AJ....153..217C}
{Cauley}, P.~W., {Redfield}, S., \& {Jensen}, A.~G. 2017, \aj, 153, 217

\bibitem[{Cherubim {et~al.}(2023)Cherubim, Cloutier, Charbonneau, Stockdale,
  Stassun, Schwarz, Safonov, Mortier, Lewin, Latham, Horne, Haywood, Gonzales,
  Goliguzova, Collins, Ciardi, Bieryla, Belinski, Wohler, Watson, Vanderspek,
  Udry, Sozzetti, Ségransan, Sasselov, Ricker, Rice, Poretti, Piotto, Pepe,
  Molinari, Micela, Mayor, Lovis, López-Morales, Jenkins, Essack, Dumusque,
  Doty, Colón, Cameron, \& Buchhave}]{Cherubim_2023}
Cherubim, C., Cloutier, R., Charbonneau, D., {et~al.} 2023, The Astronomical
  Journal, 165, 167

\bibitem[{{Choi} {et~al.}(2016){Choi}, {Dotter}, {Conroy}, {Cantiello},
  {Paxton}, \& {Johnson}}]{2016ApJ...823..102C}
{Choi}, J., {Dotter}, A., {Conroy}, C., {et~al.} 2016, \apj, 823, 102

\bibitem[{{Ciardi} {et~al.}(2015){Ciardi}, {Beichman}, {Horch}, \&
  {Howell}}]{ciardi2015}
{Ciardi}, D.~R., {Beichman}, C.~A., {Horch}, E.~P., \& {Howell}, S.~B. 2015,
  \apj, 805, 16

\bibitem[{{Collins}(2019)}]{collins:2019}
{Collins}, K. 2019, in American Astronomical Society Meeting Abstracts, Vol.
  233, American Astronomical Society Meeting Abstracts \#233, 140.05

\bibitem[{{Collins} {et~al.}(2017){Collins}, {Kielkopf}, {Stassun}, \&
  {Hessman}}]{Collins:2017}
{Collins}, K.~A., {Kielkopf}, J.~F., {Stassun}, K.~G., \& {Hessman}, F.~V.
  2017, \aj, 153, 77

\bibitem[{{Cubillos} {et~al.}(2017){Cubillos}, {Erkaev}, {Juvan}, {Fossati},
  {Johnstone}, {Lammer}, {Lendl}, {Odert}, \& {Kislyakova}}]{Cubillos2017}
{Cubillos}, P., {Erkaev}, N.~V., {Juvan}, I., {et~al.} 2017, \mnras, 466, 1868

\bibitem[{{da Silva} {et~al.}(2006){da Silva}, {Girardi}, {Pasquini},
  {Setiawan}, {von der L{\"u}he}, {de Medeiros}, {Hatzes}, {D{\"o}llinger}, \&
  {Weiss}}]{daSilva2006}
{da Silva}, L., {Girardi}, L., {Pasquini}, L., {et~al.} 2006, \aap, 458, 609

\bibitem[{{Damasso} {et~al.}(2020){Damasso}, {Sozzetti}, {Lovis}, {Barros},
  {Sousa}, {Demangeon}, {Faria}, {Lillo-Box}, {Cristiani}, {Pepe}, {Rebolo},
  {Santos}, {Zapatero Osorio}, {Gonz{\'a}lez Hern{\'a}ndez}, {Amate},
  {Pasquini}, {Zerbi}, {Adibekyan}, {Abreu}, {Affolter}, {Alibert}, {Aliverti},
  {Allart}, {Allende Prieto}, {{\'A}lvarez}, {Alves}, {Avila}, {Baldini},
  {Bandy}, {Benz}, {Bianco}, {Borsa}, {Bossini}, {Bourrier}, {Bouchy}, {Broeg},
  {Cabral}, {Calderone}, {Cirami}, {Coelho}, {Conconi}, {Coretti}, {Cumani},
  {Cupani}, {D'Odorico}, {Deiries}, {Dekker}, {Delabre}, {Di Marcantonio},
  {Dumusque}, {Ehrenreich}, {Figueira}, {Fragoso}, {Genolet}, {Genoni},
  {G{\'e}nova Santos}, {Hughes}, {Iwert}, {Kerber}, {Knudstrup}, {Landoni},
  {Lavie}, {Lizon}, {Lo Curto}, {Maire}, {Martins}, {M{\'e}gevand}, {Mehner},
  {Micela}, {Modigliani}, {Molaro}, {Monteiro}, {Monteiro}, {Moschetti},
  {Mueller}, {Murphy}, {Nunes}, {Oggioni}, {Oliveira}, {Oshagh}, {Pall{\'e}},
  {Pariani}, {Poretti}, {Rasilla}, {Rebord{\~a}o}, {Redaelli}, {Riva}, {Santana
  Tschudi}, {Santin}, {Santos}, {S{\'e}gransan}, {Schmidt}, {Segovia},
  {Sosnowska}, {Span{\`o}}, {Su{\'a}rez Mascare{\~n}o}, {Tabernero}, {Tenegi},
  {Udry}, \& {Zanutta}}]{2020A&A...642A..31D}
{Damasso}, M., {Sozzetti}, A., {Lovis}, C., {et~al.} 2020, \aap, 642, A31

\bibitem[{{Diamond-Lowe} {et~al.}(2022){Diamond-Lowe}, {Kreidberg}, {Harman},
  {Kempton}, {Rogers}, {Joyce}, {Eastman}, {King}, {Kopparapu}, {Youngblood},
  {Kosiarek}, {Livingston}, {Hardegree-Ullman}, \&
  {Crossfield}}]{2022AJ....164..172D}
{Diamond-Lowe}, H., {Kreidberg}, L., {Harman}, C.~E., {et~al.} 2022, \aj, 164,
  172

\bibitem[{Director {et~al.}(2017)Director, Gattiker, Lawrence, \&
  Wiel}]{Director17}
Director, H.~M., Gattiker, J., Lawrence, E., \& Wiel, S.~V. 2017, Journal of
  Statistical Computation and Simulation, 87, 3521

\bibitem[{{Doyle} {et~al.}(2014){Doyle}, {Davies}, {Smalley}, {Chaplin}, \&
  {Elsworth}}]{Doyle2014}
{Doyle}, A.~P., {Davies}, G.~R., {Smalley}, B., {Chaplin}, W.~J., \&
  {Elsworth}, Y. 2014, \mnras, 444, 3592

\bibitem[{{Erkaev} {et~al.}(2007){Erkaev}, {Kulikov}, {Lammer}, {Selsis},
  {Langmayr}, {Jaritz}, \& {Biernat}}]{Er07}
{Erkaev}, N.~V., {Kulikov}, Y.~N., {Lammer}, H., {et~al.} 2007, \aap, 472, 329

\bibitem[{Foreman-Mackey {et~al.}(2014)Foreman-Mackey, Hoyer, Bernhard, \&
  Angus}]{Mackey2014}
Foreman-Mackey, D., Hoyer, S., Bernhard, J., \& Angus, R. 2014, george: George
  (v0.2.0)

\bibitem[{{Fossati} {et~al.}(2017){Fossati}, {Erkaev}, {Lammer}, {Cubillos},
  {Odert}, {Juvan}, {Kislyakova}, {Lendl}, {Kubyshkina}, \&
  {Bauer}}]{2017A&A...598A..90F}
{Fossati}, L., {Erkaev}, N.~V., {Lammer}, H., {et~al.} 2017, \aap, 598, A90

\bibitem[{{Fulton} {et~al.}(2017){Fulton}, {Petigura}, {Howard}, {Isaacson},
  {Marcy}, {Cargile}, {Hebb}, {Weiss}, {Johnson}, {Morton}, {Sinukoff},
  {Crossfield}, \& {Hirsch}}]{2017AJ....154..109F}
{Fulton}, B.~J., {Petigura}, E.~A., {Howard}, A.~W., {et~al.} 2017, \aj, 154,
  109

\bibitem[{{Gandolfi} {et~al.}(2018){Gandolfi}, {Barrag{\'a}n}, {Livingston},
  {Fridlund}, {Justesen}, {Redfield}, {Fossati}, {Mathur}, {Grziwa}, {Cabrera},
  {Garc{\'{\i}}a}, {Persson}, {Van Eylen}, {Hatzes}, {Hidalgo}, {Albrecht},
  {Bugnet}, {Cochran}, {Csizmadia}, {Deeg}, {Eigm{\"u}ller}, {Endl}, {Erikson},
  {Esposito}, {Guenther}, {Korth}, {Luque}, {Monta{\~n}es Rodr{\'{\i}}guez},
  {Nespral}, {Nowak}, {P{\"a}tzold}, \& {Prieto-Arranz}}]{2018A&A...619L..10G}
{Gandolfi}, D., {Barrag{\'a}n}, O., {Livingston}, J.~H., {et~al.} 2018, \aap,
  619, L10

\bibitem[{{Garc{\'\i}a Mu{\~n}oz} {et~al.}(2021){Garc{\'\i}a Mu{\~n}oz},
  {Fossati}, {Youngblood}, {Nettelmann}, {Gandolfi}, {Cabrera}, \&
  {Rauer}}]{2021ApJ...907L..36G}
{Garc{\'\i}a Mu{\~n}oz}, A., {Fossati}, L., {Youngblood}, A., {et~al.} 2021,
  \apjl, 907, L36

\bibitem[{{Garc{\'\i}a Mu{\~n}oz} {et~al.}(2020){Garc{\'\i}a Mu{\~n}oz},
  {Youngblood}, {Fossati}, {Gandolfi}, {Cabrera}, \&
  {Rauer}}]{2020ApJ...888L..21G}
{Garc{\'\i}a Mu{\~n}oz}, A., {Youngblood}, A., {Fossati}, L., {et~al.} 2020,
  \apjl, 888, L21

\bibitem[{{Georgieva} {et~al.}(2021){Georgieva}, {Persson}, {Barrag{\'a}n},
  {Nowak}, {Fridlund}, {Locci}, {Palle}, {Luque}, {Carleo}, {Gandolfi}, {Kane},
  {Korth}, {Stassun}, {Livingston}, {Matthews}, {Collins}, {Howell}, {Serrano},
  {Albrecht}, {Bieryla}, {Brasseur}, {Ciardi}, {Cochran}, {Colon},
  {Crossfield}, {Csizmadia}, {Deeg}, {Esposito}, {Furlan}, {Gan}, {Goffo},
  {Gonzales}, {Grziwa}, {Guenther}, {Guerra}, {Hirano}, {Jenkins}, {Jensen},
  {Kab{\'a}th}, {Knudstrup}, {Lam}, {Latham}, {Levine}, {Matson}, {McDermott},
  {Osborne}, {Paegert}, {Quinn}, {Redfield}, {Ricker}, {Schlieder}, {Scott},
  {Seager}, {Smith}, {Tenenbaum}, {Twicken}, {Vanderspek}, {Van Eylen}, \&
  {Winn}}]{Geo21}
{Georgieva}, I.~Y., {Persson}, C.~M., {Barrag{\'a}n}, O., {et~al.} 2021,
  \mnras, 505, 4684

\bibitem[{{Ginzburg} {et~al.}(2018){Ginzburg}, {Schlichting}, \&
  {Sari}}]{2018MNRAS.476..759G}
{Ginzburg}, S., {Schlichting}, H.~E., \& {Sari}, R. 2018, \mnras, 476, 759

\bibitem[{{Gray}(2005)}]{Gray_book05}
{Gray}, D.~F. 2005, {The Observation and Analysis of Stellar Photospheres}

\bibitem[{{Gregory}(2005)}]{2005ApJ...631.1198G}
{Gregory}, P.~C. 2005, \apj, 631, 1198

\bibitem[{{Grimm} {et~al.}(2018){Grimm}, {Demory}, {Gillon}, {Dorn}, {Agol},
  {Burdanov}, {Delrez}, {Sestovic}, {Triaud}, {Turbet}, {Bolmont}, {Caldas},
  {de Wit}, {Jehin}, {Leconte}, {Raymond}, {Van Grootel}, {Burgasser}, {Carey},
  {Fabrycky}, {Heng}, {Hernandez}, {Ingalls}, {Lederer}, {Selsis}, \&
  {Queloz}}]{Grimm18}
{Grimm}, S.~L., {Demory}, B.-O., {Gillon}, M., {et~al.} 2018, \aap, 613, A68

\bibitem[{{Guerrero} {et~al.}(2021){Guerrero}, {Seager}, {Huang}, {Vanderburg},
  {Garcia Soto}, {Mireles}, {Hesse}, {Fong}, {Glidden}, {Shporer}, {Latham},
  {Collins}, {Quinn}, {Burt}, {Dragomir}, {Crossfield}, {Vanderspek},
  {Fausnaugh}, {Burke}, {Ricker}, {Daylan}, {Essack}, {G{\"u}nther}, {Osborn},
  {Pepper}, {Rowden}, {Sha}, {Villanueva}, {Yahalomi}, {Yu}, {Ballard},
  {Batalha}, {Berardo}, {Chontos}, {Dittmann}, {Esquerdo}, {Mikal-Evans},
  {Jayaraman}, {Krishnamurthy}, {Louie}, {Mehrle}, {Niraula}, {Rackham},
  {Rodriguez}, {Rowden}, {Sousa-Silva}, {Watanabe}, {Wong}, {Zhan},
  {Zivanovic}, {Christiansen}, {Ciardi}, {Swain}, {Lund}, {Mullally},
  {Fleming}, {Rodriguez}, {Boyd}, {Quintana}, {Barclay}, {Col{\'o}n},
  {Rinehart}, {Schlieder}, {Clampin}, {Jenkins}, {Twicken}, {Caldwell},
  {Coughlin}, {Henze}, {Lissauer}, {Morris}, {Rose}, {Smith}, {Tenenbaum},
  {Ting}, {Wohler}, {Bakos}, {Bean}, {Berta-Thompson}, {Bieryla}, {Bouma},
  {Buchhave}, {Butler}, {Charbonneau}, {Doty}, {Ge}, {Holman}, {Howard},
  {Kaltenegger}, {Kane}, {Kjeldsen}, {Kreidberg}, {Lin}, {Minsky}, {Narita},
  {Paegert}, {P{\'a}l}, {Palle}, {Sasselov}, {Spencer}, {Sozzetti}, {Stassun},
  {Torres}, {Udry}, \& {Winn}}]{Guerrero2021}
{Guerrero}, N.~M., {Seager}, S., {Huang}, C.~X., {et~al.} 2021, \apjs, 254, 39

\bibitem[{{Gupta} \& {Schlichting}(2019)}]{2019MNRAS.487...24G}
{Gupta}, A. \& {Schlichting}, H.~E. 2019, \mnras, 487, 24

\bibitem[{{Hatzes}(2019)}]{2019dmde.book.....H}
{Hatzes}, A.~P. 2019, {The Doppler Method for the Detection of Exoplanets}

\bibitem[{{Hatzes} {et~al.}(2022){Hatzes}, {Gandolfi}, {Korth}, {Rodler},
  {Sabotta}, {Esposito}, {Barrag{\'a}n}, {Van Eylen}, {Livingston}, {Serrano},
  {Luque}, {Smith}, {Redfield}, {Persson}, {P{\"a}tzold}, {Palle}, {Nowak},
  {Osborne}, {Narita}, {Mathur}, {Lam}, {Kab{\'a}th}, {Johnson}, {Guenther},
  {Grziwa}, {Goffo}, {Fridlund}, {Endl}, {Deeg}, {Csizmadia}, {Cochran},
  {Cuesta}, {Chaturvedi}, {Carleo}, {Cabrera}, {Beck}, \&
  {Albrecht}}]{2022AJ....163..223H}
{Hatzes}, A.~P., {Gandolfi}, D., {Korth}, J., {et~al.} 2022, \aj, 163, 223

\bibitem[{{Howell} {et~al.}(2011){Howell}, {Everett}, {Sherry}, {Horch}, \&
  {Ciardi}}]{howell2011}
{Howell}, S.~B., {Everett}, M.~E., {Sherry}, W., {Horch}, E., \& {Ciardi},
  D.~R. 2011, \aj, 142, 19

\bibitem[{{Howell} {et~al.}(2021){Howell}, {Matson}, {Ciardi}, {Everett},
  {Livingston}, {Scott}, {Horch}, \& {Winn}}]{2021AJ....161..164H}
{Howell}, S.~B., {Matson}, R.~A., {Ciardi}, D.~R., {et~al.} 2021, \aj, 161, 164

\bibitem[{{Howell} {et~al.}(2014){Howell}, {Sobeck}, {Haas}, {Still},
  {Barclay}, {Mullally}, {Troeltzsch}, {Aigrain}, {Bryson}, {Caldwell},
  {Chaplin}, {Cochran}, {Huber}, {Marcy}, {Miglio}, {Najita}, {Smith},
  {Twicken}, \& {Fortney}}]{2014PASP..126..398H}
{Howell}, S.~B., {Sobeck}, C., {Haas}, M., {et~al.} 2014, \pasp, 126, 398

\bibitem[{{Hu} {et~al.}(2015){Hu}, {Seager}, \& {Yung}}]{2015ApJ...807....8H}
{Hu}, R., {Seager}, S., \& {Yung}, Y.~L. 2015, \apj, 807, 8

\bibitem[{{Huang} {et~al.}(2018){Huang}, {Burt}, {Vanderburg}, {G{\"u}nther},
  {Shporer}, {Dittmann}, {Winn}, {Wittenmyer}, {Sha}, {Kane}, {Ricker},
  {Vanderspek}, {Latham}, {Seager}, {Jenkins}, {Caldwell}, {Collins},
  {Guerrero}, {Smith}, {Quinn}, {Udry}, {Pepe}, {Bouchy}, {S{\'e}gransan},
  {Lovis}, {Ehrenreich}, {Marmier}, {Mayor}, {Wohler}, {Haworth}, {Morgan},
  {Fausnaugh}, {Ciardi}, {Christiansen}, {Charbonneau}, {Dragomir}, {Deming},
  {Glidden}, {Levine}, {McCullough}, {Yu}, {Narita}, {Nguyen}, {Morton},
  {Pepper}, {P{\'a}l}, {Rodriguez}, {Stassun}, {Torres}, {Sozzetti}, {Doty},
  {Christensen-Dalsgaard}, {Laughlin}, {Clampin}, {Bean}, {Buchhave}, {Bakos},
  {Sato}, {Ida}, {Kaltenegger}, {Palle}, {Sasselov}, {Butler}, {Lissauer},
  {Ge}, \& {Rinehart}}]{2018ApJ...868L..39H}
{Huang}, C.~X., {Burt}, J., {Vanderburg}, A., {et~al.} 2018, \apjl, 868, L39

\bibitem[{{Husser} {et~al.}(2013){Husser}, {Wende-von Berg}, {Dreizler},
  {Homeier}, {Reiners}, {Barman}, \& {Hauschildt}}]{2013A&A...553A...6H}
{Husser}, T.~O., {Wende-von Berg}, S., {Dreizler}, S., {et~al.} 2013, \aap,
  553, A6

\bibitem[{{Ito} \& {Ikoma}(2021)}]{2021MNRAS.502..750I}
{Ito}, Y. \& {Ikoma}, M. 2021, \mnras, 502, 750

\bibitem[{{Jenkins}(2002)}]{Jenkins02}
{Jenkins}, J.~M. 2002, \apj, 575, 493

\bibitem[{{Jenkins} {et~al.}(2010){Jenkins}, {Chandrasekaran}, {McCauliff},
  {Caldwell}, {Tenenbaum}, {Li}, {Klaus}, {Cote}, \& {Middour}}]{Jenkins10}
{Jenkins}, J.~M., {Chandrasekaran}, H., {McCauliff}, S.~D., {et~al.} 2010, in
  Society of Photo-Optical Instrumentation Engineers (SPIE) Conference Series,
  Vol. 7740, Software and Cyberinfrastructure for Astronomy, ed. N.~M.
  {Radziwill} \& A.~{Bridger}, 77400D

\bibitem[{{Jenkins} {et~al.}(2020){Jenkins}, {Tenenbaum}, {Seader}, {Burke},
  {McCauliff}, {Smith}, {Twicken}, \& {Chandrasekaran}}]{Jenkins2020}
{Jenkins}, J.~M., {Tenenbaum}, P., {Seader}, S., {et~al.} 2020, {Kepler Data
  Processing Handbook: Transiting Planet Search}, Kepler Science Document
  KSCI-19081-003, id. 9. Edited by Jon M. Jenkins.

\bibitem[{{Jenkins} {et~al.}(2016){Jenkins}, {Twicken}, {McCauliff},
  {Campbell}, {Sanderfer}, {Lung}, {Mansouri-Samani}, {Girouard}, {Tenenbaum},
  {Klaus}, {Smith}, {Caldwell}, {Chacon}, {Henze}, {Heiges}, {Latham},
  {Morgan}, {Swade}, {Rinehart}, \& {Vanderspek}}]{jenkins2016}
{Jenkins}, J.~M., {Twicken}, J.~D., {McCauliff}, S., {et~al.} 2016, in Society
  of Photo-Optical Instrumentation Engineers (SPIE) Conference Series, Vol.
  9913, \procspie, 99133E

\bibitem[{{Jensen} {et~al.}(2012){Jensen}, {Redfield}, {Endl}, {Cochran},
  {Koesterke}, \& {Barman}}]{Jensen12}
{Jensen}, A.~G., {Redfield}, S., {Endl}, M., {et~al.} 2012, \apj, 751, 86

\bibitem[{{Kempton} {et~al.}(2018){Kempton}, {Bean}, {Louie}, {Deming}, {Koll},
  {Mansfield}, {Christiansen}, {L{\'o}pez-Morales}, {Swain}, {Zellem},
  {Ballard}, {Barclay}, {Barstow}, {Batalha}, {Beatty}, {Berta-Thompson},
  {Birkby}, {Buchhave}, {Charbonneau}, {Cowan}, {Crossfield}, {de Val-Borro},
  {Doyon}, {Dragomir}, {Gaidos}, {Heng}, {Hu}, {Kane}, {Kreidberg}, {Mallonn},
  {Morley}, {Narita}, {Nascimbeni}, {Pall{\'e}}, {Quintana}, {Rauscher},
  {Seager}, {Shkolnik}, {Sing}, {Sozzetti}, {Stassun}, {Valenti}, \& {von
  Essen}}]{Kempton18}
{Kempton}, E. M.~R., {Bean}, J.~L., {Louie}, D.~R., {et~al.} 2018, \pasp, 130,
  114401

\bibitem[{{Kipping}(2013)}]{Kipping2013}
{Kipping}, D.~M. 2013, \mnras, 435, 2152

\bibitem[{{Kite} \& {Barnett}(2020)}]{KiteandBarnett2020pnas}
{Kite}, E.~S. \& {Barnett}, M.~N. 2020, Proceedings of the National Academy of
  Science, 117, 18264

\bibitem[{Krissansen-Totton \& Fortney(2022)}]{Krissansen-Totton_2022}
Krissansen-Totton, J. \& Fortney, J.~J. 2022, The Astrophysical Journal, 933,
  115

\bibitem[{{Kubyshkina} \& {Fossati}(2022)}]{2022A&A...668A.178K}
{Kubyshkina}, D. \& {Fossati}, L. 2022, \aap, 668, A178

\bibitem[{{Kuerster} {et~al.}(1997){Kuerster}, {Schmitt}, {Cutispoto}, \&
  {Dennerl}}]{1997A&A...320..831K}
{Kuerster}, M., {Schmitt}, J.~H.~M.~M., {Cutispoto}, G., \& {Dennerl}, K. 1997,
  \aap, 320, 831

\bibitem[{{Kurucz}(1993)}]{1993yCat.6039....0K}
{Kurucz}, R.~L. 1993, VizieR Online Data Catalog, VI/39

\bibitem[{{Kurucz}(2013)}]{Kurucz2013}
{Kurucz}, R.~L. 2013, {ATLAS12: Opacity sampling model atmosphere program},
  Astrophysics Source Code Library

\bibitem[{{Lester} {et~al.}(2021){Lester}, {Matson}, {Howell}, {Furlan},
  {Gnilka}, {Scott}, {Ciardi}, {Everett}, {Hartman}, \&
  {Hirsch}}]{2021AJ....162...75L}
{Lester}, K.~V., {Matson}, R.~A., {Howell}, S.~B., {et~al.} 2021, \aj, 162, 75

\bibitem[{{Li} {et~al.}(2019){Li}, {Tenenbaum}, {Twicken}, {Burke}, {Jenkins},
  {Quintana}, {Rowe}, \& {Seader}}]{Li2019}
{Li}, J., {Tenenbaum}, P., {Twicken}, J.~D., {et~al.} 2019, \pasp, 131, 024506

\bibitem[{{Lopez} \& {Fortney}(2013)}]{2013ApJ...776....2L}
{Lopez}, E.~D. \& {Fortney}, J.~J. 2013, \apj, 776, 2

\bibitem[{{Lopez} \& {Fortney}(2014)}]{LopFor14}
{Lopez}, E.~D. \& {Fortney}, J.~J. 2014, \apj, 792, 1

\bibitem[{{Lovis} \& {Pepe}(2007)}]{2007A&A...468.1115L}
{Lovis}, C. \& {Pepe}, F. 2007, \aap, 468, 1115

\bibitem[{{Luque} \& {Pall{\'e}}(2022)}]{Luque22}
{Luque}, R. \& {Pall{\'e}}, E. 2022, Science, 377, 1211

\bibitem[{{Madhusudhan} {et~al.}(2021){Madhusudhan}, {Piette}, \&
  {Constantinou}}]{Madhu21}
{Madhusudhan}, N., {Piette}, A. A.~A., \& {Constantinou}, S. 2021, \apj, 918, 1

\bibitem[{{Malsky} {et~al.}(2022){Malsky}, {Rogers}, {Kempton}, \&
  {Marounina}}]{2022NatAs.tmp..248M}
{Malsky}, I., {Rogers}, L., {Kempton}, E. M.~R., \& {Marounina}, N. 2022,
  Nature Astronomy

\bibitem[{{Mandel} \& {Agol}(2002)}]{Mandel2002}
{Mandel}, K. \& {Agol}, E. 2002, \apjl, 580, L171

\bibitem[{{Mayor} {et~al.}(2003){Mayor}, {Pepe}, {Queloz}, {Bouchy},
  {Rupprecht}, {Lo Curto}, {Avila}, {Benz}, {Bertaux}, {Bonfils}, {Dall},
  {Dekker}, {Delabre}, {Eckert}, {Fleury}, {Gilliotte}, {Gojak}, {Guzman},
  {Kohler}, {Lizon}, {Longinotti}, {Lovis}, {Megevand}, {Pasquini}, {Reyes},
  {Sivan}, {Sosnowska}, {Soto}, {Udry}, {van Kesteren}, {Weber}, \&
  {Weilenmann}}]{Mayor03}
{Mayor}, M., {Pepe}, F., {Queloz}, D., {et~al.} 2003, The Messenger, 114, 20

\bibitem[{{McCully} {et~al.}(2018){McCully}, {Volgenau}, {Harbeck}, {Lister},
  {Saunders}, {Turner}, {Siiverd}, \& {Bowman}}]{McCully:2018}
{McCully}, C., {Volgenau}, N.~H., {Harbeck}, D.-R., {et~al.} 2018, in Society
  of Photo-Optical Instrumentation Engineers (SPIE) Conference Series, Vol.
  10707, \procspie, 107070K

\bibitem[{{Morris} {et~al.}(2020){Morris}, {Twicken}, {Smith}, {Clarke},
  {Jenkins}, {Bryson}, {Girouard}, \& {Klaus}}]{Morris20}
{Morris}, R.~L., {Twicken}, J.~D., {Smith}, J.~C., {et~al.} 2020, {Kepler Data
  Processing Handbook: Photometric Analysis}, Kepler Science Document
  KSCI-19081-003, id. 6. Edited by Jon M. Jenkins.

\bibitem[{Mousis {et~al.}(2020)Mousis, Deleuil, Aguichine, Marcq, Naar,
  Aguirre, Brugger, \& Gon{\c{c}}alves}]{Mousis20}
Mousis, O., Deleuil, M., Aguichine, A., {et~al.} 2020, The Astrophysical
  Journal, 896, L22

\bibitem[{{Owen} \& {Jackson}(2012)}]{Owen2012}
{Owen}, J.~E. \& {Jackson}, A.~P. 2012, \mnras, 425, 2931

\bibitem[{{Owen} \& {Wu}(2013)}]{2013ApJ...775..105O}
{Owen}, J.~E. \& {Wu}, Y. 2013, \apj, 775, 105

\bibitem[{{Owen} \& {Wu}(2016)}]{2016ApJ...817..107O}
{Owen}, J.~E. \& {Wu}, Y. 2016, \apj, 817, 107

\bibitem[{Parviainen(2015)}]{Parviainen2015}
Parviainen, H. 2015, MNRAS, 450, 3233

\bibitem[{{Persson} {et~al.}(2018){Persson}, {Fridlund}, {Barrag{\'a}n}, {Dai},
  {Gandolfi}, {Hatzes}, {Hirano}, {Grziwa}, {Korth}, {Prieto-Arranz},
  {Fossati}, {Van Eylen}, {Justesen}, {Livingston}, {Kubyshkina}, {Deeg},
  {Guenther}, {Nowak}, {Cabrera}, {Eigm{\"u}ller}, {Csizmadia}, {Smith},
  {Erikson}, {Albrecht}, {Sobrino}, {Cochran}, {Endl}, {Esposito}, {Fukui},
  {Heeren}, {Hidalgo}, {Hjorth}, {Kuzuhara}, {Narita}, {Nespral}, {Palle},
  {P{\"a}tzold}, {Rauer}, {Rodler}, \& {Winn}}]{2018A&A...618A..33P}
{Persson}, C.~M., {Fridlund}, M., {Barrag{\'a}n}, O., {et~al.} 2018, \aap, 618,
  A33

\bibitem[{{Persson} {et~al.}(2022){Persson}, {Georgieva}, {Gandolfi}, {Acuna},
  {Aguichine}, {Muresan}, {Guenther}, {Livingston}, {Collins}, {Dai},
  {Fridlund}, {Goffo}, {Jenkins}, {Kab{\'a}th}, {Korth}, {Levine}, {Serrano},
  {Vines}, {Barragan}, {Carleo}, {Colon}, {Cochran}, {Christiansen}, {Deeg},
  {Deleuil}, {Dragomir}, {Esposito}, {Gan}, {Grziwa}, {Hatzes}, {Hesse},
  {Horne}, {Jenkins}, {Kielkopf}, {Klagyivik}, {Lam}, {Latham}, {Luque},
  {Orell-Miquel}, {Mortier}, {Mousis}, {Narita}, {Osborne}, {Palle}, {Papini},
  {Ricker}, {Schmerling}, {Seager}, {Stassun}, {Van Eylen}, {Vanderspek},
  {Wang}, {Winn}, {Wohler}, {Zambelli}, \& {Ziegler}}]{Carina22}
{Persson}, C.~M., {Georgieva}, I.~Y., {Gandolfi}, D., {et~al.} 2022, \aap, 666,
  A184

\bibitem[{{Petigura} {et~al.}(2022){Petigura}, {Rogers}, {Isaacson}, {Owen},
  {Kraus}, {Winn}, {MacDougall}, {Howard}, {Fulton}, {Kosiarek}, {Weiss},
  {Behmard}, \& {Blunt}}]{2022AJ....163..179P}
{Petigura}, E.~A., {Rogers}, J.~G., {Isaacson}, H., {et~al.} 2022, \aj, 163,
  179

\bibitem[{{Piaulet} {et~al.}(2023){Piaulet}, {Benneke}, {Almenara}, {Dragomir},
  {Knutson}, {Thorngren}, {Peterson}, {Crossfield}, {Kempton}, {Kubyshkina},
  {Howard}, {Angus}, {Isaacson}, {Weiss}, {Beichman}, {Fortney}, {Fossati},
  {Lammer}, {McCullough}, {Morley}, \& {Wong}}]{2023NatAs...7..206P}
{Piaulet}, C., {Benneke}, B., {Almenara}, J.~M., {et~al.} 2023, Nature
  Astronomy, 7, 206

\bibitem[{{Piskunov} \& {Valenti}(2017)}]{pv2017}
{Piskunov}, N. \& {Valenti}, J.~A. 2017, \aap, 597, A16

\bibitem[{{Plotnykov} \& {Valencia}(2020)}]{Plotnykov20}
{Plotnykov}, M. \& {Valencia}, D. 2020, \mnras, 499, 932

\bibitem[{{Rajpaul} {et~al.}(2015){Rajpaul}, {Aigrain}, {Osborne}, {Reece}, \&
  {Roberts}}]{Rajpaul2015}
{Rajpaul}, V., {Aigrain}, S., {Osborne}, M.~A., {Reece}, S., \& {Roberts}, S.
  2015, \mnras, 452, 2269

\bibitem[{{Ricker} {et~al.}(2015){Ricker}, {Winn}, {Vanderspek}, {Latham},
  {Bakos}, {Bean}, {Berta-Thompson}, {Brown}, {Buchhave}, {Butler}, {Butler},
  {Chaplin}, {Charbonneau}, {Christensen-Dalsgaard}, {Clampin}, {Deming},
  {Doty}, {De Lee}, {Dressing}, {Dunham}, {Endl}, {Fressin}, {Ge}, {Henning},
  {Holman}, {Howard}, {Ida}, {Jenkins}, {Jernigan}, {Johnson}, {Kaltenegger},
  {Kawai}, {Kjeldsen}, {Laughlin}, {Levine}, {Lin}, {Lissauer}, {MacQueen},
  {Marcy}, {McCullough}, {Morton}, {Narita}, {Paegert}, {Palle}, {Pepe},
  {Pepper}, {Quirrenbach}, {Rinehart}, {Sasselov}, {Sato}, {Seager},
  {Sozzetti}, {Stassun}, {Sullivan}, {Szentgyorgyi}, {Torres}, {Udry}, \&
  {Villasenor}}]{2015JATIS...1a4003R}
{Ricker}, G.~R., {Winn}, J.~N., {Vanderspek}, R., {et~al.} 2015, Journal of
  Astronomical Telescopes, Instruments, and Systems, 1, 014003

\bibitem[{{Rogers} {et~al.}(2021){Rogers}, {Gupta}, {Owen}, \&
  {Schlichting}}]{2021MNRAS.508.5886R}
{Rogers}, J.~G., {Gupta}, A., {Owen}, J.~E., \& {Schlichting}, H.~E. 2021,
  \mnras, 508, 5886

\bibitem[{{Rogers} {et~al.}(2023){Rogers}, {Schlichting}, \&
  {Owen}}]{Rogers2023-ApJL}
{Rogers}, J.~G., {Schlichting}, H.~E., \& {Owen}, J.~E. 2023, arXiv e-prints,
  arXiv:2301.04321

\bibitem[{{Ryabchikova} {et~al.}(2015){Ryabchikova}, {Piskunov}, {Kurucz},
  {Stempels}, {Heiter}, {Pakhomov}, \& {Barklem}}]{Ryabchikova2015}
{Ryabchikova}, T., {Piskunov}, N., {Kurucz}, R.~L., {et~al.} 2015, \physscr,
  90, 054005

\bibitem[{{Sanz-Forcada} {et~al.}(2011){Sanz-Forcada}, {Micela}, {Ribas},
  {Pollock}, {Eiroa}, {Velasco}, {Solano}, \&
  {Garc{\'\i}a-{\'A}lvarez}}]{Sanz2011}
{Sanz-Forcada}, J., {Micela}, G., {Ribas}, I., {et~al.} 2011, \aap, 532, A6

\bibitem[{{Schlegel} {et~al.}(1998){Schlegel}, {Finkbeiner}, \&
  {Davis}}]{1998ApJ...500..525S}
{Schlegel}, D.~J., {Finkbeiner}, D.~P., \& {Davis}, M. 1998, \apj, 500, 525

\bibitem[{{Scott} {et~al.}(2021){Scott}, {Howell}, {Gnilka}, {Stephens},
  {Salinas}, {Matson}, {Furlan}, {Horch}, {Everett}, {Ciardi}, {Mills}, \&
  {Quigley}}]{2021FrASS...8..138S}
{Scott}, N.~J., {Howell}, S.~B., {Gnilka}, C.~L., {et~al.} 2021, Frontiers in
  Astronomy and Space Sciences, 8, 138

\bibitem[{{Smith} {et~al.}(2012){Smith}, {Stumpe}, {Van Cleve}, {Jenkins},
  {Barclay}, {Fanelli}, {Girouard}, {Kolodziejczak}, {McCauliff}, {Morris}, \&
  {Twicken}}]{Smith2012}
{Smith}, J.~C., {Stumpe}, M.~C., {Van Cleve}, J.~E., {et~al.} 2012, \pasp, 124,
  1000

\bibitem[{{Sotin} {et~al.}(2007){Sotin}, {Grasset}, \& {Mocquet}}]{Sotin07}
{Sotin}, C., {Grasset}, O., \& {Mocquet}, A. 2007, \icarus, 191, 337

\bibitem[{{Stumpe} {et~al.}(2014){Stumpe}, {Smith}, {Catanzarite}, {Van Cleve},
  {Jenkins}, {Twicken}, \& {Girouard}}]{Stumpe2014}
{Stumpe}, M.~C., {Smith}, J.~C., {Catanzarite}, J.~H., {et~al.} 2014, \pasp,
  126, 100

\bibitem[{{Stumpe} {et~al.}(2012){Stumpe}, {Smith}, {Van Cleve}, {Twicken},
  {Barclay}, {Fanelli}, {Girouard}, {Jenkins}, {Kolodziejczak}, {McCauliff}, \&
  {Morris}}]{Stumpe2012}
{Stumpe}, M.~C., {Smith}, J.~C., {Van Cleve}, J.~E., {et~al.} 2012, \pasp, 124,
  985

\bibitem[{{Tian} \& {Heng}(2023)}]{TianandHeng2023}
{Tian}, M. \& {Heng}, K. 2023, arXiv e-prints, arXiv:2301.10217

\bibitem[{{Twicken} {et~al.}(2018){Twicken}, {Catanzarite}, {Clarke},
  {Girouard}, {Jenkins}, {Klaus}, {Li}, {McCauliff}, {Seader}, {Tenenbaum},
  {Wohler}, {Bryson}, {Burke}, {Caldwell}, {Haas}, {Henze}, \&
  {Sanderfer}}]{Twicken2018}
{Twicken}, J.~D., {Catanzarite}, J.~H., {Clarke}, B.~D., {et~al.} 2018, \pasp,
  130, 064502

\bibitem[{{Twicken} {et~al.}(2010){Twicken}, {Clarke}, {Bryson}, {Tenenbaum},
  {Wu}, {Jenkins}, {Girouard}, \& {Klaus}}]{Twicken10}
{Twicken}, J.~D., {Clarke}, B.~D., {Bryson}, S.~T., {et~al.} 2010, in Society
  of Photo-Optical Instrumentation Engineers (SPIE) Conference Series, Vol.
  7740, Software and Cyberinfrastructure for Astronomy, ed. N.~M. {Radziwill}
  \& A.~{Bridger}, 774023

\bibitem[{{Valencia} {et~al.}(2007){Valencia}, {Sasselov}, \&
  {O'Connell}}]{Valencia07}
{Valencia}, D., {Sasselov}, D.~D., \& {O'Connell}, R.~J. 2007, \apj, 665, 1413

\bibitem[{{Valenti} \& {Piskunov}(1996)}]{vp96}
{Valenti}, J.~A. \& {Piskunov}, N. 1996, \aaps, 118, 595

\bibitem[{{Van Eylen} {et~al.}(2018){Van Eylen}, {Agentoft}, {Lundkvist},
  {Kjeldsen}, {Owen}, {Fulton}, {Petigura}, \& {Snellen}}]{VanEylen2018}
{Van Eylen}, V., {Agentoft}, C., {Lundkvist}, M.~S., {et~al.} 2018, \mnras,
  479, 4786

\bibitem[{{Vines} \& {Jenkins}(2022)}]{2022MNRAS.513.2719V}
{Vines}, J.~I. \& {Jenkins}, J.~S. 2022, \mnras, 513, 2719

\bibitem[{{Wildi} {et~al.}(2010){Wildi}, {Pepe}, {Chazelas}, {Lo Curto}, \&
  {Lovis}}]{Wildi10}
{Wildi}, F., {Pepe}, F., {Chazelas}, B., {Lo Curto}, G., \& {Lovis}, C. 2010,
  in Society of Photo-Optical Instrumentation Engineers (SPIE) Conference
  Series, Vol. 7735, Ground-based and Airborne Instrumentation for Astronomy
  III, ed. I.~S. {McLean}, S.~K. {Ramsay}, \& H.~{Takami}, 77354X

\bibitem[{{Wildi} {et~al.}(2011){Wildi}, {Pepe}, {Chazelas}, {Lo Curto}, \&
  {Lovis}}]{Wildi11}
{Wildi}, F., {Pepe}, F., {Chazelas}, B., {Lo Curto}, G., \& {Lovis}, C. 2011,
  in Society of Photo-Optical Instrumentation Engineers (SPIE) Conference
  Series, Vol. 8151, Techniques and Instrumentation for Detection of Exoplanets
  V, ed. S.~{Shaklan}, 81511F

\bibitem[{{Winn}(2010)}]{2010exop.book...55W}
{Winn}, J.~N. 2010, {Exoplanet Transits and Occultations}, ed. S.~{Seager}
  (University of Arizona Press), 55--77

\bibitem[{{Yee} {et~al.}(2017){Yee}, {Petigura}, \& {von
  Braun}}]{2017ApJ...836...77Y}
{Yee}, S.~W., {Petigura}, E.~A., \& {von Braun}, K. 2017, \apj, 836, 77

\bibitem[{{Zechmeister} \& {K{\"u}rster}(2009)}]{Zech09}
{Zechmeister}, M. \& {K{\"u}rster}, M. 2009, \aap, 496, 577

\bibitem[{{Zechmeister} {et~al.}(2018){Zechmeister}, {Reiners}, {Amado},
  {Azzaro}, {Bauer}, {B{\'e}jar}, {Caballero}, {Guenther}, {Hagen}, {Jeffers},
  {Kaminski}, {K{\"u}rster}, {Launhardt}, {Montes}, {Morales}, {Quirrenbach},
  {Reffert}, {Ribas}, {Seifert}, {Tal-Or}, \& {Wolthoff}}]{Zechmeister2018}
{Zechmeister}, M., {Reiners}, A., {Amado}, P.~J., {et~al.} 2018, \aap, 609, A12

\bibitem[{{Zeng} {et~al.}(2021){Zeng}, {Jacobsen}, {Hyung}, {Levi}, {Nava},
  {Kirk}, {Piaulet}, {Lacedelli}, {Sasselov}, {Petaev}, {Stewart}, {Alam},
  {L{\'o}pez-Morales}, {Damasso}, \& {Latham}}]{2021ApJ...923..247Z}
{Zeng}, L., {Jacobsen}, S.~B., {Hyung}, E., {et~al.} 2021, \apj, 923, 247

\bibitem[{{Zeng} {et~al.}(2019){Zeng}, {Jacobsen}, {Sasselov}, {Petaev},
  {Vanderburg}, {Lopez-Morales}, {Perez-Mercader}, {Mattsson}, {Li}, {Heising},
  {Bonomo}, {Damasso}, {Berger}, {Cao}, {Levi}, \& {Wordsworth}}]{Zeng2019}
{Zeng}, L., {Jacobsen}, S.~B., {Sasselov}, D.~D., {et~al.} 2019, Proceedings of
  the National Academy of Science, 116, 9723

\bibitem[{{Zeng} {et~al.}(2016){Zeng}, {Sasselov}, \& {Jacobsen}}]{Zeng2016}
{Zeng}, L., {Sasselov}, D.~D., \& {Jacobsen}, S.~B. 2016, \apj, 819, 127

\end{thebibliography}

\begin{appendix}
\section{HARPS data}
\null\newpage
\clearpage
\begin{sidewaystable}
\begin{tiny}
\begin{center}
  \caption{Radial velocities and spectral activity indicators measured from 3.6m/HARPS spectra. Full data available in machine-readable format as supplementary material.
    \label{all_rv.tex}}
  \begin{tabular}{rrrrrrrrrrrrrrrrrrrrr}
    \hline
    \hline
    \multicolumn{1}{r}{BJD$_\mathrm{TDB}$ (days)} &
    \multicolumn{1}{r}{RV} &
    \multicolumn{1}{r}{$\sigma_\mathrm{RV}$} &
    \multicolumn{1}{r}{dlW} &
    \multicolumn{1}{r}{$\sigma_\mathrm{dlW}$} &
    \multicolumn{1}{r}{crx} &
    \multicolumn{1}{r}{$\sigma_\mathrm{crx}$} &
    \multicolumn{1}{r}{$\mathrm{H_{\alpha}}$} &
    \multicolumn{1}{r}{BIS} &
    \multicolumn{1}{r}{CCF\_FHWM} &
    \multicolumn{1}{r}{S-index} &
    \multicolumn{1}{r}{$\sigma_\mathrm{S-index}$} &
    \multicolumn{1}{r}{SNR} &
    \multicolumn{1}{r}{$\mathrm{T_{exp}}$}\\ 
    \multicolumn{1}{r}{---} &
    \multicolumn{1}{r}{($\mathrm{km\,s^{-1}}$)} &
    \multicolumn{1}{r}{($\mathrm{km\,s^{-1}}$)} &
    \multicolumn{1}{r}{($\mathrm{km^{-2}\,s^{-2}}$)} &
    \multicolumn{1}{r}{($\mathrm{km^{-2}\,s^{-2}}$} &
    \multicolumn{1}{r}{($\mathrm{m\,s^{-1}}$)} &
    \multicolumn{1}{r}{($\mathrm{m\,s^{-1}}$)} &
    \multicolumn{1}{r}{---} &
    \multicolumn{1}{r}{($\mathrm{km\,s^{-1}}$)} &
    \multicolumn{1}{r}{($\mathrm{km\,s^{-1}}$)} &
    \multicolumn{1}{r}{---} &
    \multicolumn{1}{r}{---} &
    \multicolumn{1}{r}{---} &
    \multicolumn{1}{r}{(s)}\\ 
    \hline
     2459626.74 & -23.81634664 & 0.000848944 & 3.846147994 & 1.326126148 & 13.66794426 & 10.57927121 & 0.983948301 & -0.016951275 & 6.954078739 & 0.181637014 & 0.001014454 & 99.8 & 1500 \\ 
        2459627.78 & -23.81252261 & 0.000959742 & 2.075952242 & 1.43990305 & 17.52283094 & 10.32374355 & 0.982740024 & -0.024461918 & 6.957723772 & 0.185009486 & 0.001177316 & 89.6 & 1700 \\ 
        2459628.784 & -23.80894807 & 0.000679691 & 5.772780144 & 1.201660819 & -6.274365888 & 6.13518358 & 0.981689555 & -0.016765188 & 6.960223013 & 0.18452779 & 0.000862516 & 123.9 & 1500 \\ 
        2459628.825 & -23.81073021 & 0.000736545 & 5.175949702 & 1.522926677 & -15.7972952 & 8.05105209 & 0.984447113 & -0.019061706 & 6.960107901 & 0.184778148 & 0.000996763 & 118.3 & 1500 \\ 
        2459629.796 & -23.81694989 & 0.000745206 & 9.040799319 & 1.206887028 & 8.160118161 & 9.015379128 & 0.985988722 & -0.014237337 & 6.965701117 & 0.183937567 & 0.000967848 & 114.9 & 1500 \\ 
        2459630.815 & -23.8240658 & 0.00089424 & 7.293146335 & 1.311379023 & -4.711933434 & 8.969100789 & 0.983616792 & -0.013837771 & 6.960550131 & 0.184412118 & 0.001190777 & 98.7 & 1500 \\ 
        2459631.753 & -23.82450219 & 0.00090718 & 3.520022825 & 1.506126036 & -20.84466988 & 8.441737559 & 0.980787294 & -0.01223902 & 6.959490586 & 0.1856573 & 0.001113644 & 94.1 & 1500 \\ 
        2459631.849 & -23.82561963 & 0.000842201 & 4.591938152 & 1.457629472 & -3.851899306 & 8.621181622 & 0.981372138 & -0.014852202 & 6.965124773 & 0.1874519 & 0.001137536 & 104.9 & 1800 \\ 
        2459633.762 & -23.8179495 & 0.001008271 & -0.243779863 & 1.493026406 & -10.68601408 & 10.15031443 & 0.978736037 & -0.013287999 & 6.952109623 & 0.187487834 & 0.001233629 & 85.4 & 1500 \\ 
        2459633.817 & -23.8194531 & 0.000985333 & 0.967864549 & 1.437758574 & -12.3525101 & 9.455692577 & 0.984426883 & -0.016698413 & 6.957616563 & 0.186071554 & 0.00125368 & 89.2 & 1800 \\ 
        2459634.654 & -23.81874425 & 0.000929297 & -0.437742299 & 1.647382169 & -0.204005127 & 7.581201672 & 0.985609597 & -0.016231192 & 6.955020371 & 0.182129452 & 0.001089646 & 91.3 & 1500 \\ 
        ...  \\
    \hline
  \end{tabular}
\end{center}
\end{tiny}
\end{sidewaystable}

\section{Frequency analysis of the FWHM residuals}

\begin{figure}[!ht]
\centering
\includegraphics[width=1\linewidth]{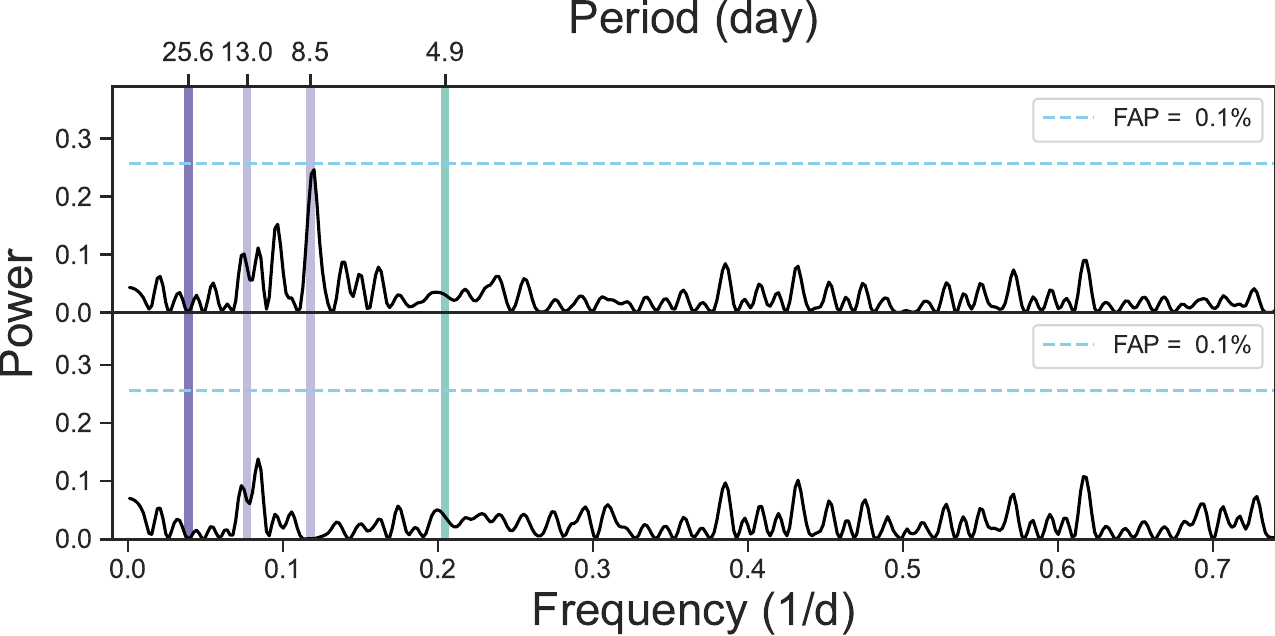}
\caption{GLS periodogram of the FWHM residuals after subtracting the 25.6-day signal (Fig.\ref{fig: GLS}, panel d;}   top panel). The 8.5-day signal seen in the RV panels of Fig.\ref{fig: GLS} becomes significant. The bottom panel shows the FWHM after subtracting both the 25.6-, and the 8.5-day signals. Here, the 12.8-day signal remains, but is not significant.
\label{Figure: GLS_res}
\end{figure}

\end{appendix}
 
\end{document}